\documentclass[table]{interact}
\usepackage{natbib}
\bibpunct[, ]{(}{)}{,}{a}{}{,}
\renewcommand\bibfont{\fontsize{10}{12}\selectfont}
\usepackage{todonotes,soul}
\usepackage{paralist,enumerate,tabularx}
\usepackage{subcaption}
\usepackage{url}
\usepackage{tabularx}
\usepackage{color}
\usepackage{tabularray}
\definecolor{Silver}{rgb}{0.752,0.752,0.752}
\usepackage{multirow}

\begin{document}
\articletype{Research Paper}
\title{The KnowWhereGraph: A Large-Scale Geo-Knowledge Graph for Interdisciplinary Knowledge Discovery and Geo-Enrichment}
\author{
\name{
Rui Zhu\textsuperscript{a}\thanks{CONTACT Rui Zhu Email: rui.zhu@bristol.ac.uk}, 
Cogan Shimizu\textsuperscript{b},
Shirly Stephen\textsuperscript{c}, 
Colby K. Fisher\textsuperscript{d},
Thomas Thelen\textsuperscript{e},
Kitty Currier\textsuperscript{c},
Krzysztof Janowicz\textsuperscript{f,c}, 
Pascal Hitzler\textsuperscript{g},
Mark Schildhauer\textsuperscript{c},
Wenwen Li\textsuperscript{h}, 
Dean Rehberger\textsuperscript{i},
Adrita Barua\textsuperscript{g},
Antrea Christou\textsuperscript{b},
Ling Cai\textsuperscript{c},
Abhilekha Dalal\textsuperscript{g},
Anthony D'Onofrio\textsuperscript{i},
Andrew Eells\textsuperscript{g},
Mitchell Faulk\textsuperscript{c},
Zilong Liu\textsuperscript{f,c},
Gengchen Mai\textsuperscript{j},
Mohammad Saeid Mahdavinejad\textsuperscript{g},
Bryce Mecum\textsuperscript{c},
Sanaz Saki Norouzi\textsuperscript{g},
Meilin Shi\textsuperscript{f},
Yuanyuan Tian\textsuperscript{h},
Sizhe Wang\textsuperscript{h},
Zhangyu Wang\textsuperscript{c}, 
Joseph Zalewski\textsuperscript{g}
}
\affil{
\textsuperscript{a}University of Bristol, Bristol, UK; 
\textsuperscript{b}Wright State University, OH, U.S.;
\textsuperscript{c}University of California, Santa Barbara, CA, U.S.;
\textsuperscript{d}Hydronos Labs, NJ, U.S.;
\textsuperscript{e}Independent Scholar, CA, U.S.;
\textsuperscript{f}University of Vienna, Vienna, Austria;
\textsuperscript{g}Kansas State University, KS, U.S.;
\textsuperscript{h}Arizona State University, AZ, U.S.;
\textsuperscript{i}Michigan State University, MI, U.S.;
\textsuperscript{j}University of Texas at Austin, TX, U.S.
}
}
\maketitle
\begin{abstract}
  Global challenges such as food supply chain disruptions, public health crises, and natural hazard responses require access to and integration of diverse datasets, many of which are geospatial. Over the past few years, a growing number of (geo)portals have been developed to address this need. However, most existing (geo)portals are stacked by separated or sparsely connected data ``silos" impeding effective data consolidation. A new way of sharing and reusing geospatial data is therefore urgently needed. In this work, we introduce KnowWhereGraph, a knowledge graph-based data integration, enrichment, and synthesis framework that not only includes schemas and data related to human and environmental systems but also provides a suite of supporting tools for accessing this information. The KnowWhereGraph aims to address the challenge of data integration by building a large-scale, cross-domain, pre-integrated, FAIR-principles-based, and AI-ready data warehouse rooted in knowledge graphs. We highlight the design principles of KnowWhereGraph, emphasizing the roles of space, place, and time in bridging various data ``silos". Additionally, we demonstrate multiple use cases where the proposed geospatial knowledge graph and its associated tools empower decision-makers to uncover insights that are often hidden within complex and poorly interoperable datasets.
\end{abstract}
\section{Introduction}
\label{sec:intro}
Knowledge graphs have become an established and actively growing framework for representing, retrieving, and integrating data from diverse and heterogeneous sources \citep{hogan2021knowledge,hitzler-cacm}. Recently, knowledge graphs, together with their associated tools, have played key roles in the development of AI-powered search engines, personal assistants, and decision support systems \citep{kgs}. However, their success in managing and analyzing human and environmental observations remains limited, despite the abundance of large-scale geospatial data.  

To meet this need, we have developed KnowWhereGraph, one of the world's largest publicly accessible geospatial knowledge graphs~\citep{janowicz2022know}. 
KnowWhereGraph is implemented using the Resource Description Framework (RDF) and encompasses \textbf{over 29 billion RDF triples} (i.e., graph statements), representing densely-integrated statements across diverse domains such as disasters, demographics, soils, health, and so on.
KnowWhereGraph is designed to support a wide range of applications, including those in the humanitarian relief, food, and agriculture sectors, with a focus on supply chain management. It is also used to tackle key environmental policy challenges, such as agricultural sustainability, soil conservation, land valuation and risk assessment, as well as humanitarian aid allocation in response to natural disasters domestically and internationally. 
To achieve this, KnowWhereGraph integrates over 30 datasets, covering a spectrum of topics including environment observations (e.g., soil type, climate observations, air pollution), natural hazards and their impacts (e.g., storms, drought, and floods), demographic and public health metrics (e.g., poverty status, social vulnerability, medical infrastructures), and humanitarian relief-related data (e.g., local experts and their expertises).

This large-scale data integration aims at answering specific queries, which we refer to as \emph{area briefings}. These briefings deliver up-to-date, contextualized, i.e., geo-enriched, information about human and environmental factors within seconds, enabling decision-makers and data scientists to gain rapid situational awareness. The scope of these briefings covers several key questions, including general descriptions of the area (e.g., What is here, and what are its characteristics?), historical events (e.g., What has happened here in the past?), experts with knowledge about the area (e.g., Who has expertise on this region?), and relationships between different areas (e.g., How is the wildfire in Area A related to the flood in Area B?). These questions are addressed from multiple perspectives, accounting for different conceptualizations of geographic regions, such as ZIP code areas and climate divisions. 


The diverse and critical use cases, the vast array of data sources, and the integration of various interlinked regional identifiers presented unique challenges throughout the project's life-cycle, such as difficulties in conflating and querying both raster and vector data, inconsistency of spatial and temporal resolutions across data sources, as well as ambiguous semantics of the data attributes and provenance. Addressing these challenges require the development of innovative approaches to structuring the knowledge graph, ensuring efficient querying and enhanced usability. 
To tackle these complexities, KnowWhereGraph utilizes a rigorously structured, modular schema designed to systematically enhance interoperability across diverse geographic datasets. This schema serves to structure and constrain the graph, ensuring efficient data integration, retrieval, and application. 

\begin{figure}[ht]
    \centering
    \includegraphics[width=\textwidth]{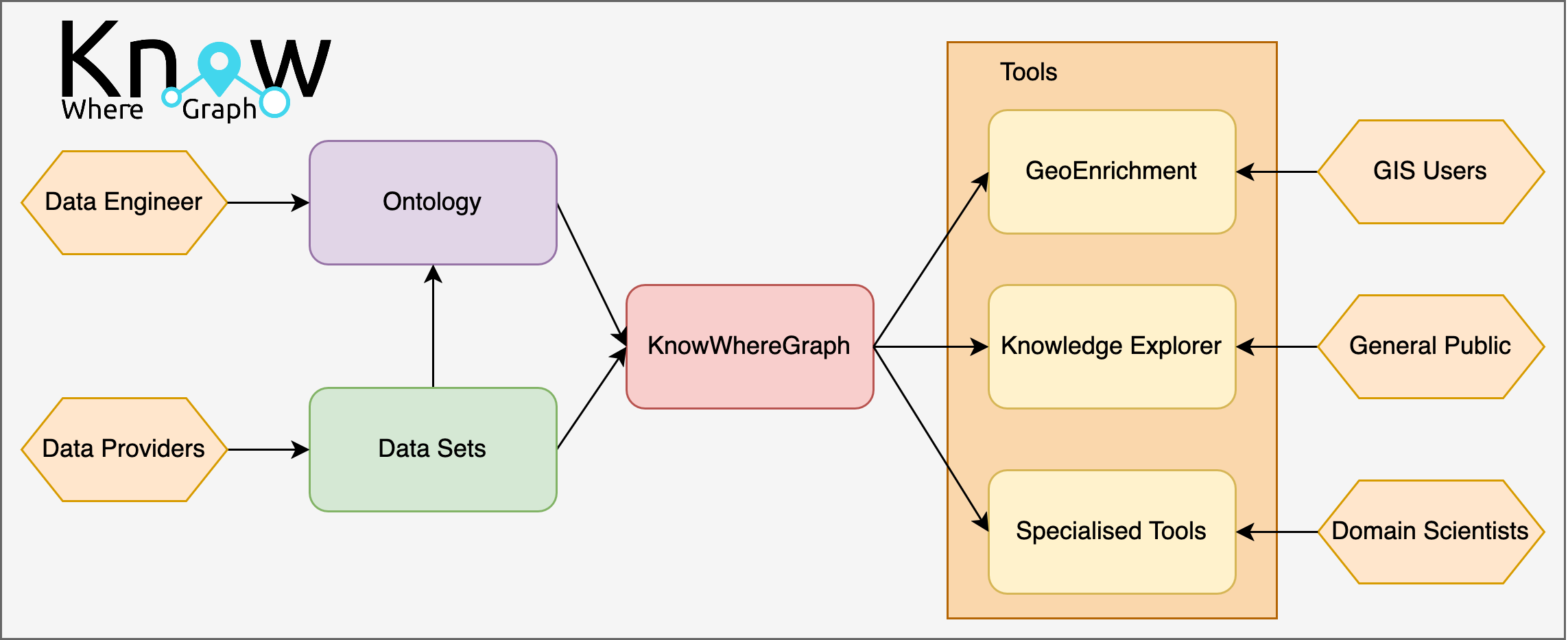}
    \caption{Overview structure of KnowWhereGraph.}
    \label{fig:KnowWhereGraph-overview}
\end{figure}

\textbf{The contributions of this paper are as follows:}
\begin{compactitem}
    \item We introduce the key design philosophy underlying KnowWhereGraph;
    \item We discuss the curated data collection that forms the foundation of KnowWhereGraph;
    \item We introduce the modular ontology serving as the schema for structuring the graph and run queries against it; 
    \item and we present the KnowWhereGraph ecosystem of supporting tools and services.
\end{compactitem}
These elements are visually represented in Figure~\ref{fig:KnowWhereGraph-overview}. Surrounding the core component of this work---KnowWhereGraph---are the datasets, ontology, and tools. Essentially, KnowWhereGraph is built by first identifying a set of use cases and data sources that are of interest to our partners, especially those in the sectors of humanitarian relief, the food supply chain, and agriculture. Subsequently, data engineers co-design the ontology (referred to as data scheme in this paper) with these partners based on a variety of considerations, including the nature of the data, existing data standards and best practices from the community, use case demands, and querying performance. To enable a wide use of KnowWhereGraph, we develop a series of tools for users from multiple disciplines with various levels of technical skill in GIS and knowledge graph--related technologies. For example, geoenrichment tools are designed specifically for those who are familiar with GIS software such as ArcGIS Desktop or QGIS. Knowledge Explorer is implemented as an interactive map-based Web interface for the general public. To support domain scientists, we have also developed customized tools to access KnowWhereGraph and conduct bespoke analysis so as to inform decision-making, such as identifying supply chain disruptions resulting from a wildfire, assessing land potentials to plan for the growing of different kinds of crops, and providing medical resources in response to hurricanes. The list of datasets, designed schema, and codes of implementing these tools are shared in our public Github repository\footnote{\url{https://github.com/KnowWhereGraph}} and we provide a project webpage\footnote{\url{https://knowwheregraph.org/}} as an ultimate venue to access all these resources too. As part of NSF's Convergence Accelerator program, KnowWhereGraph was co-designed with domain experts and industry partners deeply embedded into the project's team.

The remainder of this paper is structured as follows.
The following section discusses how KnowWhereGraph advances the current state-of-the-art.
Section~\ref{sec:data} details the datasets integrated into KnowWhereGraph.
In Section~\ref{sec:schema}, we review the schema underlying KnowWhereGraph and its various modules, linked vocabularies, and ontology network.
Section~\ref{sec:tools} describes the bespoke tools developed to support various use cases enabled by KnowWhereGraph. 
Finally, Section~\ref{sec:conc} presents our conclusions.

\section{Related Work}
\label{sec:rel}
In this section, we discuss the role of spatial data infrastructures and (geo)portals in facilitating the development and deployment of geospatial knowledge graphs. We focus on how existing geospatial knowledge graphs compare with KnowWhereGraph. Finally, we discuss several key concepts, methodologies, and conventions used in this paper.  
\subsection{Spatial Data Infrastructures}
\label{ssec:sdi}
Central to discussions of geospatial data is the concept of Spatial Data Infrastructure (SDI), which encompasses the fundamental physical and organizational structures that enable the use of spatial data by multiple stakeholders \citep{rajabifard_role_2006}. Typical definitions invoke some combination of components---e.g., technology, standards, systems, people, policies, and data---aiming at facilitating data acquisition, storage, and sharing, assisting policy-making, and enhancing business operations \citep{hendriks_reconsidering_2012}. The five most typical activities performed within SDIs are identified as data discovery, access, registration, processing, and visualization. However, these activities often encounter semantic challenges, such as missing or unclear metadata, which can stem from both the original data and the processes involved \citep{janowicz2010semantic}. To overcome these challenges and extend the capabilities of SDIs, Linked Data, a term synonymous with knowledge graphs, has been proposed to semantically empower SDIs. Linked Data is increasingly recognized as the cornerstone of next-generation SDIs \citep{janowicz2010semantic, wiemann2016spatial, huang2019assessment, ronzhin2019next, huang2020towards}. KnowWhereGraph, presented in this paper, is one such initiative aiming at building a cross-disciplinary SDI using Linked Data technology. 

\subsection{(Geo)portals}
\label{ssec:gp}
The key activities identified within SDIs have driven the development of (geo)portals, defining the essential tasks these platforms are expected to perform. (Geo)portals are web-based geospatial resources that enable searching and accessing integrated geographic information \citep{maguire2005emergence,mai2020semantically}. By following the Open Geospatial Consortium (OGC) and the International Organization for Standardization (ISO) standards, (geo)portals have achieved functionalities such as metadata cataloging, data discovery, data visualization, data sharing, and data downloading. To address the challenge of topic heterogeneity resulting from multiple metadata standards, \citet{hu2015metadata} proposed Linked Data--driven (geo)portals with enhanced interoperability. Furthermore, \citet{jiang2019current} provided a summary and highlighted remaining challenges in existing (geo)portals, such as data updating, harmonization, and the need to address multidisciplinary demands in the era of big data.

(Geo)portals remain crucial for data sharing, visualization, and online management, especially within government agencies and research institutions. However, there is a growing need for (geo)portals to transition toward more comprehensive geospatial knowledge graphs, i,e., \textit{making data smart}, not applications \citep{janowicz2015data}. Such graphs provide more advanced analysis across domains to better manage the ever-growing volume of geospatial data. In this paper, we outline the strategies and methods employed by KnowWhereGraph to build an open and user-friendly platform. A key focus is on bridging the gap between data from different themes and sources that are typically stored separately in conventional (geo)portals.

\subsection{Geospatial Knowledge Graphs}
\label{ssec:gkg}
With the availability of large-scale encyclopedic knowledge graphs like DBpedia \citep{auer2007dbpedia} and YAGO \citep{suchanek2007yago}, there is an increasing need to link rich geospatial datasets for improved data discovery, integration, and knowledge extraction \citep{mai2022symbolic,regalia2018gnis,qi2023evkg}. Previous efforts, such as the LinkGeoData project \citep{auer2009linkedgeodata,stadler2012linkedgeodata}, have focused on transforming OpenStreetMap (OSM)\footnote{\url{https://www.openstreetmap.org/}} data into a graph structure, linking it with DBpedia, GeoNames\footnote{\url{https://www.geonames.org/}}, and other datasets. YAGO2 \citep{hoffart2013yago2} adds spatio-temporal dimensions to YAGO by including \emph{time} and \emph{location} in its original subject–property–object triples representation. YAGO2geo \citep{karalis2019extending} further extends YAGO2 by incorporating more detailed geospatial information, such as administrative data from three countries, geometries of administrative regions from the Global Administrative Areas dataset (GADM), and geographic features like lakes and streams from OSM. These advancements, for instance, representing cities with multipolygons instead of simple coordinates, significantly enhanced geospatial capabilities for more complex question-answering \citep{mai2021geographic}. However, these geospatial knowledge graphs have mainly focused on modeling and integrating a \textit{limited} set of regional identifiers, thus limiting their ability to cover a broader range of geospatial information, such as land-use delineation, climate zones, and hurricane trajectories. This paper introduces a novel combination of discrete global grid (DGG) and knowledge graph technology used by KnowWhereGraph to harmonize different types of geographic information across various themes.

Finally, it is worth noting the recent launch of ArcGIS Knowledge\footnote{\url{https://enterprise.arcgis.com/en/knowledge/}} in 2023, an enterprise knowledge graph tool that combines graph and spatial analytics. This tool enables the discovery and analysis of interlinked spatial and non-spatial, structured and unstructured data. Additionally, the latest release\footnote{\url{https://enterprise.arcgis.com/en/knowledge/latest/introduction/what-is-arcgis-knowledge.htm}} of ArcGIS Knowledge also allows users to connect existing graph databases to the ArcGIS platform. To align with these industry advancements, this paper also introduces plug-ins for both ArcGIS Pro and QGIS\footnote{\url{https://qgis.org/}}, allowing users to readily benefit from the rich and interlinked data provided by KnowWhereGraph. 


\subsection{Convention and Background}
\label{ssec:conv}
We note a few conventions that are used throughout the paper and provide background discussions surrounding key techniques adopted in this paper.

\textbf{Schema Diagrams.} These are used to quickly and intuitively convey the expected structure of a knowledge graph. They are central to the Modular Ontology Modeling (MOMo) process \citep{momo-swj}, which was the initial knowledge modeling methodology used to develop KnowWhereGraph's schema, and they serve as our primary visual tool for both team collaboration and outreach concerning symbolic knowledge models. These diagrams are consistently used throughout our documentation, and they have a stable visual syntax.

Each schema diagram represents a labeled graph that indicates Web Ontology Language (OWL) entities and their (possible) relations. Nodes in these diagrams are categorized and visually distinguished as follows: (1) \emph{Classes} -- rectangular, gold, solid border, (2) \emph{Modules} -- rectangular, light blue, dashed border, (3) \emph{Controlled vocabularies} -- rectangular, purple, solid border, (4) \emph{Data types} -- oval, yellow, solid border. Arrows in the schema diagrams can be of two types: (1) \emph{Subclass relation} (\texttt{rdfs:subClassOf}) -- white-headed without labels, (2) \emph{Data}/\emph{object property} -- labeled with the name of the property. A through discussion of these constructs, along with the ontology modeling process that guided the construction of KnowWhereGraph's underlying graph, is provided by \citet{kwg-jws,kwg-fois,momo-swj}. Such schema diagrams can be found in Figures~\ref{fig:KnowWhereGraph-core-sosa-kernel}--\ref{fig:md-model}.

\textbf{URI and Prefix.}
Table~\ref{tab:namespaces} lists the various namespaces, in the format of URI (Uniform Resource Identifier), and their prefixes that we use within KnowWhereGraph, including the datatypes and ontologies that we re-use. Of particular importance are \texttt{kwg-ont}, which is used for KnowWhereGraph's ontology, and \texttt{kwgr}, which generally indicates individual entity or instance. For example, it is common to see the triple of the following form to indicate the statement that ``x (e.g., Thomas Fire) is an instance of type X (e.g., wildfire), defined by KnowWhereGraph ontology (Section \ref{sec:schema})". 
\begin{equation*}
    \langle\texttt{kwgr:x, a, kwg-ont:X}\rangle
\end{equation*}

\begin{table}
    \centering
    \begin{tabular}{r|l}
Prefix   & Namespace \\\hline
geo:     & $\langle$\url{http://www.opengis.net/ont/geosparql#}$\rangle$\\
cdt:    & $\langle$\url{http://w3id.org/lindt/custom_datatypes#}$\rangle$\\
kwg-ont: & $\langle$\url{http://stko-kwg.geog.ucsb.edu/lod/ontology/}$\rangle$\\
kwgr:    & $\langle$\url{http://stko-kwg.geog.ucsb.edu/lod/resource/}$\rangle$\\
prov:    & $\langle$\url{http://www.w3.org/ns/prov#}$\rangle$\\
qudt:    & $\langle$\url{http://qudt.org/schema/qudt/}$\rangle$\\
sosa:    & $\langle$\url{http://www.w3.org/ns/sosa/}$\rangle$\\
time:    & $\langle$\url{http://www.w3.org/2006/time#}$\rangle$\\
xsd:     & $\langle$\url{http://www.w3.org/2001/XMLSchema#}$\rangle$
    \end{tabular}
    \caption{Namespaces used in KnowWhereGraph and in the various schema diagrams presented in this paper.} 
    \label{tab:namespaces}
\end{table}



\textbf{Discrete Global Grids. } A Discrete Global Grid (DGG) system is a framework composed of hierarchical levels of grids. The key components of a DGG include (1) \emph{cells} -- the basic unit, representing a fixed geographic region; (2) \emph{levels} -- corresponding to granularities of cells; and (3) \emph{coverings} -- a collection of cells that cover specific areas or shapes. Each cell in a DGG is uniquely identified by a stable cell ID, which encodes information about its level, relations with its parent/children, and neighboring cells from the same level. 

In KnowWhereGraph, we utilize Google's \emph{S2 Geometry} framework \citep{veach2017s2}, which is a hierarchical mosaic of spherical quadrilateral cells, as the DGG foundation. S2 cells are sequentially indexed using a Hilbert space-filling curve that traverses the surface of a unit sphere projected onto the six faces of a cube. This projection ensures a relatively even distribution of cells across the sphere's surface. Each S2 cell is identified by a unique 64-bit \texttt{S2CellID}, which encodes the cell’s location on the curve and its level within the S2 hierarchy. 

\section{Data Sources and Graph Overview}
\label{sec:data}

\subsection{Data Sources}
This section introduces various datasets selected and semantically annotated as part of KnowWhereGraph. Several criteria guided the selection process. First, to support our goal of creating an open knowledge network, only open datasets from government agencies (e.g., U.S. Department of Agriculture) or popular third-party sources (e.g., Healthcare Ready) were chosen. Second, as a pilot initiative, these datasets were selected to align with the specific needs of our use cases with partners, which are further discussed in Section \ref{sec:tools}. Finally, when data were not available from open sources, we designed methods to collect this information independently. For example, to assist our partner Direct Relief\footnote{\url{https://www.directrelief.org/}} in identifying local experts for disaster response, we developed custom approaches to gather raw data on experts and their expertise (see Section \ref{ssec:expert}). 

Table~\ref{tab:datalist} provides an overview of the data sources integrated into KnowWhereGraph, including themes such as environmental observations, natural hazards, transportation networks, public health, and humanitarian relief--related expert--expertise data. Currently, KnowWhereGraph mainly focuses on data from the U.S., although the workflow is designed to be adaptable for global applications. The temporal coverage of each dataset varies depending on its source, and we aim to update KnowWhereGraph with new datasets. To keep KnowWhereGraph consistent while evolving, meta and provenance details for each dataset, including spatial and temporal coverage, source, last update, and more, are represented within the graph through a metadata framework, which is discussed further in Section~\ref{ssec:md-model}.

\begin{table}[ht!]
\centering
\caption{List of thematic data sources. Yellow rows indicate environment observations; red for natural hazards; blue for transport network; purple for public health; and green for humanitarian relief--related expert-expertise data. Note that the spatial coverage of all these data sources is in the U.S. except the Comprehensive Earthquake Catalog and Relief Web Reports, which are global; ``current" in the Temporal Coverage column refers to the latest release of KnowWhereGraph. At the moment of drafting this paper, the latest release date is April 2024. The full name of the agency acronyms is listed in Appendix A. }
\label{tab:datalist}
\resizebox{\textwidth}{!}{%
{\renewcommand{\arraystretch}{1.2}
\begin{tabular}{|l|l|l|l|}
\hline
\rowcolor[HTML]{C0C0C0} 
\textbf{Dataset Name}                                                                                               & \textbf{\begin{tabular}[c]{@{}l@{}}Source \\ Agency\end{tabular}} & \textbf{Key Attributes}                                                                                                                                                & \textbf{\begin{tabular}[c]{@{}l@{}}Temporal \\ Coverage\end{tabular}} \\ \hline
\rowcolor[HTML]{FFFFF0} 
Soil Survey Geographic Database                                                                                     & NRCS                                                              & soil type, soil property, etc.                                                                                                                                         & 2000-current                                                          \\
\rowcolor[HTML]{FFFFF0} 
Cropland Data Layer                                                                                                 & USDA                                                              & crop type, crop property, etc.                                                                                                                                         & 2008-current                                                          \\
\rowcolor[HTML]{FFFFF0} 
Climate Observational Data                                                                                          & NOAA                                                              & \begin{tabular}[c]{@{}l@{}}temperature, precipitation,\\ drought severity index,\\ heating/cooling degree days, etc.\end{tabular}                                      & 1895-current                                                          \\
\rowcolor[HTML]{FFFFF0} 
Air Quality System Data                                                                                             & USEPA                                                             & \begin{tabular}[c]{@{}l@{}}Pb, CO, ozone, PM2.5, \\ NO2, SO2, PM10, etc.\end{tabular}                                                                                  & 1980-current                                                          \\
\rowcolor[HTML]{FFFFF0} 
BlueSky Daily Runs                                                                                                  & USFS                                                              & \begin{tabular}[c]{@{}l@{}}daily prediction of area of fires, \\ PM2.5, PM10, etc.\end{tabular}                                                                        & current                                                               \\ \hline
\rowcolor[HTML]{FFDFDD} 
\begin{tabular}[c]{@{}l@{}}Monitoring Trends in Burn \\ Severity Burned Areas Boundaries\end{tabular}               & USGS                                                              & \begin{tabular}[c]{@{}l@{}}no. acres burned, \\ wildfire risk and severity \\ (e.g., dNBR value), etc.\end{tabular}                                                    & 1984-current                                                          \\
\rowcolor[HTML]{FFDFDD} 
Wildland Fire Incident Locations                                                                                    & NIFC                                                              & no. acres burned, estimated cost, etc.                                                                                                                                 & 2014-current                                                          \\
\rowcolor[HTML]{FFDFDD} 
\begin{tabular}[c]{@{}l@{}}Hazard Mapping System Fire \\ and Smoke Product\end{tabular}                             & NOAA                                                              & smoke density, etc                                                                                                                                                     & 2005-current                                                          \\
\rowcolor[HTML]{FFDFDD} 
Comprehensive Earthquake Catalog                                                                                    & USGS                                                              & magnitude, depth, type, etc.                                                                                                                                           & 1843-current                                                          \\
\rowcolor[HTML]{FFDFDD} 
Historical Hurricane Tracks                                                                                         & NOAA                                                              & pressure, wind speed, etc.                                                                                                                                             & 1842-current                                                          \\
\rowcolor[HTML]{FFDFDD} 
Storm Events                                                                                                        & NOAA                                                              & \begin{tabular}[c]{@{}l@{}}no. injured and dead, \\ property damage, etc.\end{tabular}                                                                                 & 1950-current                                                          \\
\rowcolor[HTML]{FFDFDD} 
Drought Monitor Dataset                                                                                             & NDMC                                                              & drought intensity, etc.                                                                                                                                                & 2020-current                                                          \\
\rowcolor[HTML]{FFDFDD} 
OpenFEMA Disaster Declarations                                                                                      & FEMA                                                              & \begin{tabular}[c]{@{}l@{}}designated area, program, \\ amount approved, \\ program designated date, etc.\end{tabular}                                                 & 1953-current                                                          \\ \hline
\rowcolor[HTML]{DAE8FC} 
\begin{tabular}[c]{@{}l@{}}The National Highway Planning \\ Network Dataset\end{tabular}                            & USDOT                                                             & road type, road length, road sign, etc.                                                                                                                                & current                                                               \\ \hline
\rowcolor[HTML]{E3D9E4} 
\begin{tabular}[c]{@{}l@{}}Hospital - Homeland Infrastructure \\ Foundation-Level Data\end{tabular}                 & USDHS                                                             & \cellcolor[HTML]{E3D9E4}                                                                                                                                               & \cellcolor[HTML]{E3D9E4}                                              \\
\rowcolor[HTML]{E3D9E4} 
\begin{tabular}[c]{@{}l@{}}Public Health Departments -Homeland \\ Infrastructure Foundation-Level Data\end{tabular} & USDHS                                                             & \cellcolor[HTML]{E3D9E4}                                                                                                                                               & \cellcolor[HTML]{E3D9E4}                                              \\
\rowcolor[HTML]{E3D9E4} 
Rx Open Facilities                                                                                                  & \begin{tabular}[c]{@{}l@{}}Healthcare \\ Ready\end{tabular}       & \multirow{-5}{*}{\cellcolor[HTML]{E3D9E4}\begin{tabular}[c]{@{}l@{}}address, phone no., personnel, \\ helipad, trauma center level, \\ no. of beds, etc.\end{tabular}} & \multirow{-5}{*}{\cellcolor[HTML]{E3D9E4}current}                     \\
\rowcolor[HTML]{E3D9E4} 
Poverty and Obesity - PLACES                                                                                        & CDC                                                               & diabetes, obesity, poverty, etc.                                                                                                                                       & current                                                               \\
\rowcolor[HTML]{E3D9E4} 
Social Vulnerability Index                                                                                          & CDC                                                               & social vulnerability index, etc.                                                                                                                                       & 2020                                                                  \\
\rowcolor[HTML]{E3D9E4} 
\begin{tabular}[c]{@{}l@{}}Poverty Status - \\ Small Area Income and \\ Poverty Estimates Program\end{tabular}      & USCB                                                              & \begin{tabular}[c]{@{}l@{}}poverty household count, \\ poverty rate, etc.\end{tabular}                                                                                 & 2021                                                                  \\
\rowcolor[HTML]{E3D9E4} 
Health Rankings National Data                                                                                       & USCB                                                              & \begin{tabular}[c]{@{}l@{}}food environment index, \\ no. of males/females, \\ no. of household units, etc.\end{tabular}                                               & 2023                                                                  \\ \hline
\rowcolor[HTML]{E5FFE8} 
Federally Qualified Health Centers                                                                                  & HRSA                                                              & \begin{tabular}[c]{@{}l@{}}total patients, specialty, full-time \\ equivalent of personnel for a specialty\end{tabular}                                              & current                                                               \\
\rowcolor[HTML]{E5FFE8} 
Relief Web Reports                                                                                                  & UN OCHA                                                           & \begin{tabular}[c]{@{}l@{}}organizations, report type,\\ GLIDE no., region\end{tabular}                                                                                & 1980-current                                                          \\ \hline
\end{tabular}
}}
\end{table}

Compared to other knowledge graphs, such as Wikidata\footnote{\url{https://www.wikidata.org/wiki/Wikidata:Main_Page}}, KnowWhereGraph specializes in integrating datasets at the human--environment interface with the regional identifiers as the linking nexus. Namely, it focuses on formally representing these different forms of regional identifiers, such as ZIP code region and climate division, as well as their connections, which are mostly spatial (e.g., \textit{is adjacent to} and \textit{touches}). With cross-linked regions, thematic data, such as census statistics or climate observations, can be linked through their associated regional identifiers. Such a design principle rests on the observation that all environmental and societal events and phenomena happen at some specific locations within a given time frame. Therefore, space and time can be regarded as the common ground for integrating multi-sourced datasets. Nevertheless, harmonizing the representation of space is non-trivial, because data providers design distinct ways of discretizing space for different purposes (e.g., making sure each census region has a similar population; minimizing climate dynamics within each climate zone). This inevitably leads to a diverse set of regional identifiers to reference geographically different datasets. Table \ref{tab:regional-id} shows a collection of regional identifiers that are most often used in open datasets provided by the U.S. government or third parties, all of which are included in KnowWhereGraph. To cross-link them, we have developed an ontology that combines the existing GeoSPARQL Ontology with DGGs, which will be discussed in detail in Section \ref{ssec:st}.  

\begin{table}
\centering
\caption{List of regional identifiers and their data sources. The full names of the defining authorities are listed in Appendix~\ref{app:acronym}.}
\label{tab:regional-id}
\begin{tabular}{lll}
\hline
Dataset Name                                   & \begin{tabular}[c]{@{}l@{}}Defining\\ Authority\end{tabular} & \begin{tabular}[c]{@{}l@{}}Spatial \\ Coverage\end{tabular} \\ \hline
Global Administrative Areas                    & GADM.org                                                     & Global                                                      \\
National Weather Service Public Forecast Zones & NOAA                                                         & U.S.                                                        \\
ZIP Code Tabulation Areas                      & USCB                                                         & U.S.                                                        \\
Federal Information Processing Standards Code  & NRCS                                                         & U.S.                                                        \\
Climate Division                               & NOAA                                                         & U.S.                                                        \\
Census Metropolitan Areas                      & USCB                                                         & U.S.                                                        \\
U.S. Federal Judicial Districts                  & USDoJ                                                        & U.S.                                                        \\ \hline
\end{tabular}
\end{table}

\subsection{Graph Overview}
Due to the dynamics of the selected datasets, KnowWhereGraph keeps evolving as new data is added, the underlying schema is adjusted, external graphs are linked, and more. To keep track of these changes, we adopted versioning strategies. Generally, we release a new graph when there are major changes, such as the introduction of new regional identifiers and the addition of their cross-links with other regional identifiers in the graph. Minor updates, such as adding a statement or removing an old dataset, are only documented on our GitHub page without triggering the release of a new version. 

At the time of writing this paper, the third KnowWhereGraph public release (named ``Santa Barbara"\footnote{The prior releases ``Vienna" and ``Manhattan" contained $\approx$ 10 and 15 billion graph statements, respectively.}) was published, consisting of about 29 billion statements, of which approximately 20 billion are originally generated and the remaining 8 billion are inferred. The graph is formalized using RDF, an international standard recommended by W3C to connect data on the web. Specifically, statements are represented as a triple \texttt{$\langle$subject, predicate, object$\rangle$} in RDF. For instance, the statement ``Thomas Fire occurred in Santa Barbara" is represented as \texttt{$\langle$Thomas\_Fire, occurred\_in, Santa\_Barbara$\rangle$}. Note that each node and edge (e.g., \texttt{Thomas\_Fire} and \texttt{occurred\_in}) are referenced using a URI to increase their interoperability and reuse on the web. The graph is available via a public endpoint\footnote{\url{https://stko-kwg.geog.ucsb.edu/graphdb/}} supported by GraphDB\footnote{\url{https://graphdb.ontotext.com/}} where data can be retrieved using (Geo)SPARQL, a standard query language for RDF graphs. 

\section{The KnowWhereGraph Schema}
\label{sec:schema}
\begin{figure}[t]
    \centering
    \includegraphics[scale=0.35]{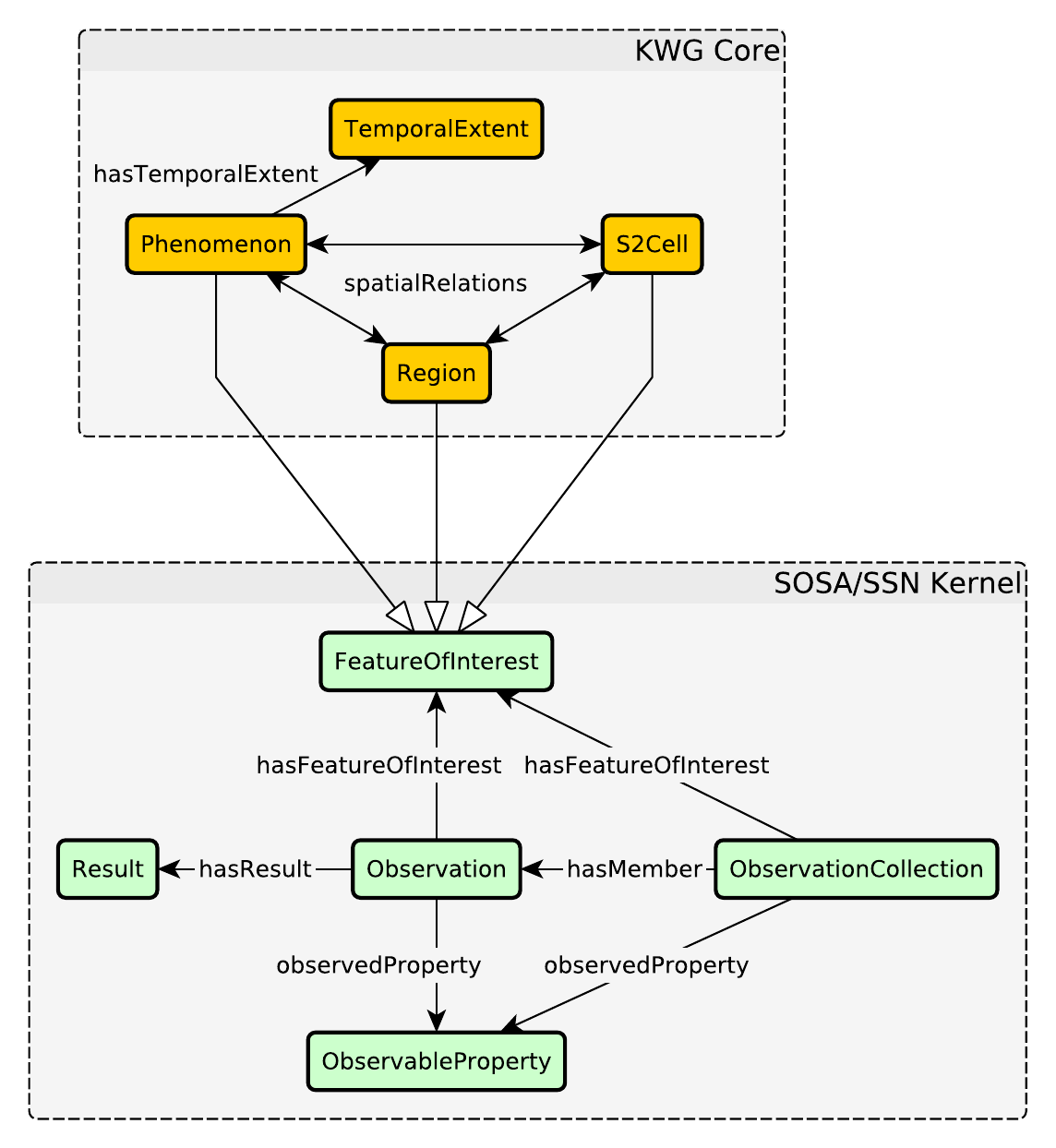}
    \caption{This figure shows the highest level of abstraction that still depicts the meaningful structure of KnowWhereGraph. Pictured on top is the geospatial backbone, coupled with the bottom, how phenomena are conceptualized and subsequently integrated with the DGG.}
    \label{fig:KnowWhereGraph-core-sosa-kernel}
\end{figure}

The KnowWhereGraph schema is rather expansive, covering: (1) Environmental observations, including measurements, units, and properties; (2) Time and space, including a model for the DGG; (3) Disasters and phenomenology; (4) Experts and their expertise; (5) Metadata about KWG itself; (6) Causal events; and (7) Alignment between scientific taxonomies. We discuss these components in the following. A high-level abstraction of KnowWhereGraph schema is depicted in Figure \ref{fig:KnowWhereGraph-core-sosa-kernel}.

\subsection{Modeling Environmental Observations}
\label{ssec:env}
\begin{figure}[ht]
    \centering
        \includegraphics[width=0.8\textwidth]{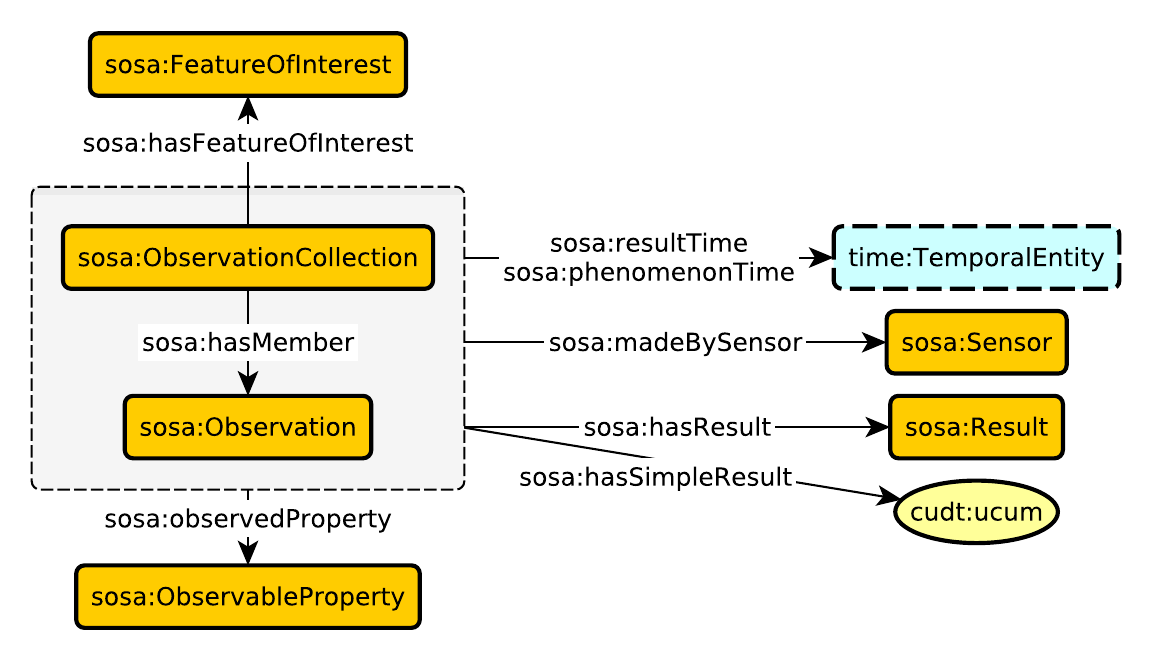}
        \caption{This schema diagram shows the classes that are reused in KnowWhereGraph from the SOSA/SSN Ontology, OWL Time Ontology, QUDT (Quantities, Units, Dimensions, and Types Ontology), and Custom Data Types ontology.}
        \label{fig:sosa-excerpt}
\end{figure}

To model environmental observations, we reuse a subset of the SOSA (Sensor, Observation, Sample, and Actuator)/SSN (Semantic Sensor Network) Ontology \citep{sosa-tr,sosa-jws,ssn-swj}. These classes and their relations are shown in Figure~\ref{fig:sosa-excerpt}. As the number of data layers integrated into KnowWhereGraph increases, this subset may be expanded. Also, to fit the nature of the raw data collected by different agencies, we flexibly adopt the SOSA/SSN ontology, making sure it precisely expresses the semantics of the data. For example, at the outset of development, we did not use the \texttt{sosa:Sensor} class since there is no data layer that describes them. This exemplifies our gradual reuse of external ontologies and vocabularies and materializes domain-specific data accordingly. While modeling environmental observations using the SOSA ontology, we particularly attempt to address four challenges: (1) Consistently represent different spatial objects and phenomena; (2) Group observations sharing the same properties; (3) Model both historical observations and future forecasts using the same ontology; and (4) Validate the modeling. 

First, in KnowWhereGraph, the \texttt{sosa:FeatureOfInterest} can be many things; primarily, \texttt{kwg:Hazard} and \texttt{kwg:Region}, two of the largest classes, can be modeled as its sub-classes. 
This allows the same structure to describe the impacts of a hazard (e.g., the magnitude of an earthquake) and public health characteristics of a region (e.g., the social vulnerability index). This facilitates data integration (i.e., the conceptual pattern remains similar even across heterogeneous datasets) and simplifies ad-hoc querying since the expected pattern of observations about the feature of interest remains identical. 

Secondly, the purpose of the \texttt{sosa:ObservationCollection} class is to provide a layer of abstraction that allows for the grouping of a particular set of \texttt{sosa:Observation} instances along different axes (e.g., collected at the same time or have the same feature of interest). This helps reduce the materialization burden but is a trade-off with query complexity. However, not all data layers will benefit from the use of a \texttt{sosa:ObservationCollection}. The trade-offs on whether or not the use of the collection is appropriate are further explored in \cite{env-kg}.

Thirdly, KnowWhereGraph contains a variety of data layers over time, including historical data (e.g., historical fires and earthquakes), near real-time data (e.g., air pollutants and drought zones), and forecasts (e.g., BlueSky smoke plumes forecast). To support these, we use the relations \texttt{sosa:resultTime} and \texttt{sosa:phenomenonTime} in a principled manner. In particular, when the \texttt{sosa:TemporalEntity} connected via \texttt{sosa:resultTime} precedes the \texttt{sosa:phenomenonTime}, we know that the effect, impact, or hazard, is a prediction or forecast. 

Finally, in KnowWhereGraph, we validate our materialization using SHACL (Shapes Constraint Language) shapes \citep{shacl-tr} and have provided a library of generic SOSA/SSN shapes: SOSA-SHACL \citep{kwg-shacl}. In general, it defines a collection of shape constraints to verify if the materialized graph complies with the SOSA recommendation and reports any violations if not. It is used to control the data quality while adding new observations to KnowWhereGraph and could be used beyond the project whenever one uses SOSA/SSN for data modeling. 

\subsection{Modeling Time and Space}
\label{ssec:st}
Observations in KnowWhereGraph are temporally and spatially indexed. For example, each smoke plume record includes both temporal information (when this smoke plume started and when it was recorded) and spatial information (the polygon that represents the extent of the smoke plume). Harmonizing the representation of temporal and spatial information is among the key goals of KnowWhereGraph. To do so, we standardize temporal and spatial reference systems in the following ways: for temporal reference, we preset temporal literal templates of multiple granularities; for spatial reference, we reuse the GeoSPARQL ontology to formalize different forms of geographic regions and adopt DGGs (e.g., S2 Cells) as the ultimate integrator for geometries.  




\subsubsection{Temporal Reference}
\begin{figure}[ht]
    \centering
    \includegraphics[width=0.5\textwidth]{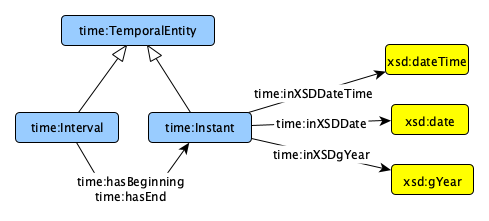}
        \caption{Time Ontology used in KnowWhereGraph, adopted from Time-OWL.}
        \label{fig:time-ont}
\end{figure}

Table \ref{tab:temporal-ref} lists the temporal templates we use to model time in KnowWhereGraph, and Figure \ref{fig:time-ont} illustrates the time model. The choice of granularity depends on the raw data. For example, the U.S. National Oceanic and Atmospheric Administration (NOAA)'s Smoke Plume dataset contains time stamps up to the \textit{Date} granularity while NOAA natural disasters record temporal granularity as \textit{Date + Time}. Note that thanks to the use of standardized Time Ontology in OWL\footnote{\url{https://www.w3.org/TR/owl-time/}}, we are able to reason over temporal information regardless of what granularity the data is associated with (e.g., \texttt{"1950-01-03"ˆˆxsd:date} is \textit{before} \texttt{"1952"ˆˆxsd:gYear}). Also, both temporal instances and intervals (with start and end time being formally defined) are applied to provide a flexible way of modeling various forms of temporal information. 

\begin{table*}[t!]
  \caption{Temporal referencing template. }
  \label{tab:temporal-ref}
  \centering \small \setlength{\tabcolsep}{3pt}
  \makebox[\textwidth]{\begin{tabular}{lll}
    \toprule
    Temporal granularity & Template & Predicate \\
    \midrule
    Date + Time & ``1950-01-03T11:00:00+06:00"\^{}\^{}xsd:dateTime & time:inXSDDateTime \\
    Date & ``1950-01-03"\^{}\^{}xsd:date & time:inXSDDate \\
    Year & ``1950"\^{}\^{}xsd:gYear & time:inXSDgYear \\
    \bottomrule
  \end{tabular}}
\end{table*}

\subsubsection{Geographic Regions}
\label{ssec:st-region}
\begin{figure}[ht]
    \centering
    \includegraphics[width=0.7\textwidth]{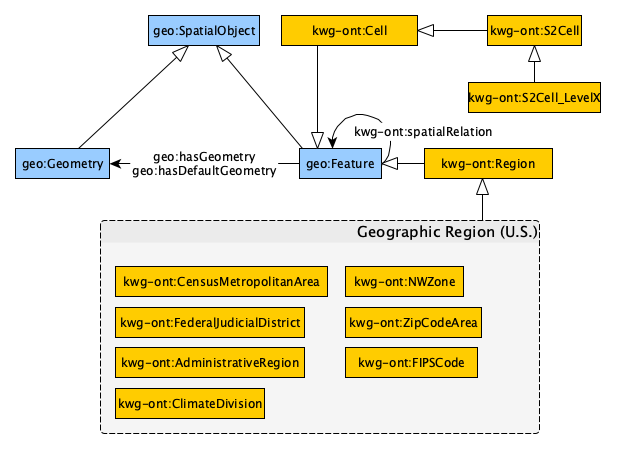}
    \caption{Spatial Ontology used in KnowWhereGraph, adopted from GeoSPARQL Ontology. Blue boxes are classes from GeoSPARQL and yellow ones are newly designed for KnowWhereGraph. The class names of geographic regions are derived from the regional identifiers listed in  Table~\ref{tab:regional-id}. Particularly, \textbf{NWZone} corresponds to ``National Weather Service Public Forecast Zones". }
    \label{fig:space-ont}
\end{figure}

The key attribute of a geospatial knowledge graph, such as KnowWhereGraph, is that most data are geographically indexed \citep{zhu2024geo-kg}. Namely,  data are recorded in relation to some locations on the surface of the Earth. Hence, KnowWhereGraph leverages identifiers, such as geographic regions, as the nexus to link datasets across the boundary of different disciplines. Nevertheless, regionalization of geographic space varies across datasets mainly due to distinct sampling strategies used to meet the specific needs of collecting the data. For instance, ZIP code regions were designed based on the delineation of administrative regions, their hierarchy, and the street network to serve postal services. In contrast, climate divisions are aggregated regions that share similar meteorological variables, such as temperature and precipitations, to provide relatively homogeneous measurements within regions. To accommodate the diversity of geographic regions used in various datasets, and more importantly, to enhance interoperability across ``siloed" datasets, KnowWhereGraph standardizes the representation of geographic regions by designing a spatial model, as depicted in Figure~\ref{fig:space-ont}. This makes KnowWhereGraph a gazetteer in the sense that it integrates different region and place identifiers and then delivers information about these identifiers, e.g., past events, demographics, etc.

We create a new class \texttt{kwg-ont:Region} as an interface for KnowWhereGraph to reuse the GeoSPARQL ontology. This class encapsulates the set of different forms of geographic regions. To further improve the linkage between datasets, we formalize spatial relations the following Region Connection Calculus (RCC8) \citep{cohn1997qualitative} while using OGC's naming system, as detailed in Table \ref{tab:spatial-ref}. Notably, since the GeoSPARQL ontology also defines spatial relations in the same way with a prefix \texttt{geo}, we define spatial relations using a distinct prefix \texttt{kwg}. It distinguishes spatial relations that are pre-computed in KnowWhereGraph (with prefix \texttt{kwg}) and those queried on-the-fly through GeoSPARQL (with prefix \texttt{geo}). The former uniquely advances the query performance of KnowWhereGraph and geographically enriches connections among datasets by explicitly materializing spatial relations between geometries of observations. 

\subsubsection{Geo-referenced Cells}\label{sec:geo-referenced-cells}
The introduced \texttt{kwg-ont:Region} only represents spatial objects that have clearly defined boundaries (i.e., vector data), while falling short of covering other forms of geospatial data, such as rasters, networks, geo-referenced text, etc. Besides, existing methods only support data integration and query via pairwise spatial relations and well-defined regions, e.g., to find hazards that have impacted (e.g., whose geometries intersected) both Santa Barbara County and Ventura County. However, it is challenging to offer a holistic situation awareness to decision-makers given an arbitrary location on the surface of the Earth, e.g., to find all hazards that happened close to Highway 101 and the variation of impacts to their local neighborhoods. 

The DGG, using the S2 grid as a specific example in this study, serves as a framework for dividing the unit sphere into a hierarchical structure of cells. Each cell is a quadrilateral defined by four geodesics. While S2 cell levels range from 0 to 30, in practice, we use levels 8 (roughly covers $1300\ km^2$) to 13 (roughly $1.3\ km^2$), as most geographic phenomena and events in KnowWhereGraph fall between these spatial scales. For each geometry of an observation, we compute its relative spatial relation with S2 cells. 

\begin{figure}[t]
    \centering
    \includegraphics[width=0.6\textwidth]{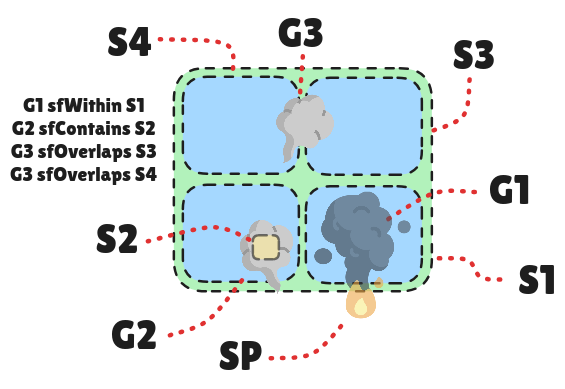}
    \caption{This smoke plume (\texttt{SP}) has three associated geometries: \texttt{G1-3}. These have spatial relations to cells: \texttt{S2-4}. Using the various spatial relations between cells allows for continuous (partitioned) coverage of the entire smoke plume.}
    \label{fig:cell-relations}
\end{figure}

For example, if an observation \texttt{SP} of a smoke plume has three geometries \texttt{G1}, \texttt{G2} and \texttt{G3}, and they have spatial relations with four cells: \texttt{S1-4}. These relations are demonstrated in Figure~\ref{fig:cell-relations}. Consider the following relations. \texttt{<G1, kwg-ont:sfWithin, S1>}, \texttt{<G2, kwg-ont:sfContains, S1>}, \texttt{<G3, kwg-ont:sfOverlaps, S3>} and \texttt{<G3, kwg-ont:sfOverlaps, S4>}, then when users query events that are related to \texttt{S1}, they will obtain information for the entire \texttt{SP}, and if they query about any other cell, since they are within \texttt{S1}, there is an intuitive mechanism to navigate through spatial granularity of impacted locations and retrieve additional context. In this way, KnowWhereGraph provides situational awareness to end users about any location on Earth's surface.

Table \ref{tab:spatial-ref} lists the relative spatial relations we model in KnowWhereGraph and their descriptions, which are borrowed from OGC's spatial predicates\footnote{\url{https://www.ogc.org/standard/sfa/}}. Among all five relations, the most frequently used three are \textbf{Within}, \textbf{Containing}, and \textbf{Overlapping} in KnowWhereGraph. Using these relations, we can answer queries like ``How many fire instances happened in the region of Santa Barbara (or any arbitrarily defined polygon)" and ``How many counties did Hurricane Laura affect in 2020". Especially, for each pair of observations and S2 cells, we materialize both the relation from the observation geometry to S2 cells and the inverse relation, i.e., the relation from S2 cells to the observation geometry. For example, if we have \texttt{<G1, kwg-ont:sfWithin, S1>} in KnowWhereGraph, we will also have \texttt{<S1, kwg-ont:sfContains, G1>}. This allows users to query S2 cells with observations, and more importantly, query observations in relation with other observations through S2 cells. That is, querying ``What are the observations that are spatially related to observation A" now can be modeled as first querying ``What are the S2 cells that are related to observation A" and then querying ``What are the observations related to the S2 cells returned from the previous query". This is the fundamental mechanism of spatial query in KnowWhereGraph, which is distinct to any other geospatial knowledge graphs.
\begin{table*}[t!]
  \caption{Spatial referencing. }
  \label{tab:spatial-ref}
  \centering \small \setlength{\tabcolsep}{3pt}
  \makebox[\textwidth]{\begin{tabular}{llll}
    \toprule
    Spatial relation & Predicate & Inverse predicate & Description \\
    \midrule
    Equality & kwg-ont:sfEquals & kwg-ont:sfEquals & \begin{tabular}{@{}l@{}}Two geometries match in both \\ type and shape.\end{tabular} \\
    Within & kwg-ont:sfWithin & kwg-ont:sfContains & \begin{tabular}{@{}l@{}}Geometry of subject is within \\ geometry of object.\end{tabular} \\
    Containing & kwg-ont:sfContains & kwg-ont:sfWithin & \begin{tabular}{@{}l@{}}Geometry of subject contains \\ geometry of object. \end{tabular} \\
    Touching & kwg-ont:sfTouches & kwg-ont:sfTouches & \begin{tabular}{@{}l@{}}Two geometries touch from outside.\end{tabular} \\
    Overlapping & kwg-ont:sfOverlaps & kwg-ont:sfOverlaps & \begin{tabular}{@{}l@{}}Two geometries partially overlap.\end{tabular} \\
    \bottomrule
  \end{tabular}}
\end{table*}

\subsection{Modeling Hazard and Its Impacts}
\label{ssec:hazard}
\begin{figure}[ht]
    \centering
    \includegraphics[width=0.8\linewidth]{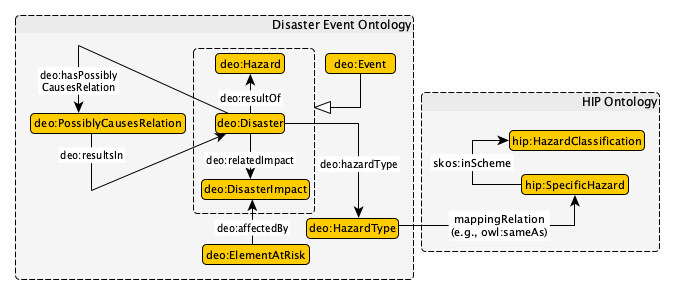}
    \caption{A high-level overview of the DEO and the HIP Ontology from \cite{hip-fois}.}
    \label{fig:disaster-ontology}
\end{figure}

Two specific ontologies were developed to model disaster-themed data within KnowWhereGraph: the Disaster Event Ontology (DEO) and the Hazard Information Profiles (HIP) Ontology \citep{hip-fois} (see an overview in Figure~\ref{fig:disaster-ontology}). DEO conceptualizes disaster events, including their impacts, classification schemes, associated properties, risk elements, and spatiotemporal characteristics. A key objective for data integration and querying in KnowWhereGraph is to establish connections across heterogeneous datasets, e.g., disaster occurrences, related observations, impacts, affected regions, and demographic information. This includes linking storm impacts with storm tracks, disaster declarations, and fire-related phenomena such as burn scars, smoke plumes, and fire forecasts.

The DEO focuses on three main classes: \texttt{deo:Event}, \texttt{deo:ElementAtRisk}, and \texttt{deo:PossiblyCausesRelation}. The observational, spatial, and temporal contexts of events are modeled using ontologies designed in Section \ref{ssec:env} and \ref{ssec:st}. The \texttt{deo:Event} class generalizes \texttt{deo:Hazard}, \texttt{deo:Disaster}, and \texttt{deo:DisasterImpact}, which are subclasses of SOSA’s \texttt{sosa:FeatureOfInterest}. By reusing standard ontologies (e.g., SOSA, GeoSPARQL ontology) at the top level, we were able to use a simple, consistent ontology pattern for uniformly representing and querying across all hazard observational data (e.g., droughts, hurricanes, wildfires) in KnowWhereGraph. The \texttt{deo:ElementAtRisk} class represents entities of value that may be adversely affected by hazards, such as demographics, public health infrastructure, and transportation facilities. Additionally, the Disaster Properties Ontology \citep{kwg-disaster} was developed to model event properties (e.g., severity of impact) and elements at risk (e.g., resilience, vulnerability), though it is not yet actively used within KnowWhereGraph. Causal relations between events are represented using the reified \texttt{deo:PossiblyCausesRelation} from the Causal ODP \citep{odp-ce}.

The HIP Ontology was developed to formalize the United Nations (UN) Office for Disaster Risk Reduction (UNDRR) Hazard Classification Scheme as a FAIR-principles-based vocabulary to align and disambiguate hazard types in KnowWhereGraph. Each hazard type class in the ontology is annotated with specific details, e.g., the hazard type name, reference number, authoritative definitions, the UN organization providing guidance, and supplementary annotations like synonyms, scientific descriptions, and metrics. Using the HIP ontology, we have integrated various disaster-themed datasets from NOAA, FEMA, and MTBS with authoritative classification schemes and vocabularies from the UNDRR (see Appendix \ref{app:acronym} for the full name of these agencies).

\subsection{Modeling Expert and Their Expertise}
\label{ssec:expert}
We developed the Expertise Ontology (ExO) \citep{kwg-expertise} in KnowWhereGraph to consistently represent a variety of expertise-related information. ExO encompasses information on people known for their expertise in humanitarian-relief-related topics, such as specific named diaster, specific types of disaster, local emergency response policy, and public health planning. ExO is built around a fundamental framework of classes and properties designed to represent experts and their areas of expertise. Essentially, we model expertise as a hierarchy of topics, so that one could query expertise in different levels of granularity, which are further linked to other nodes in the graph for enrichment. Since these topics could be either research- or experience-based through specific activities, we scoped them spatially and temporally using the scheme discussed in Section \ref{ssec:st}. 


The concept of topics within our expert ontology ranges from broad knowledge domains, such as science and nanoscience, to highly specific domains, such as the impact of Hurricane Harvey in Houston, 2017. Two distinctive datasets were used, each meticulously represented in KnowWhereGraph. The first dataset comprises academic professionals with specific expertise in the realm of disaster response. We identified these experts through an in-depth analysis of data extracted from Google Scholar and Semantic Scholar publications. Their proficiency was assessed quantitatively via the semantic similarity algorithm, with supplementary information regarding employment and affiliation acquired from organization homepages. The second dataset spans U.S. Federally Qualified Health Centers (FQHC), offering a wealth of potential expertise crucial for humanitarian aid responses. These centers harbor various affiliates such as dentists, clinicians, and other medical staff, each with their respective specialties. 

While traditional domain ontologies employ a class-subclass hierarchy, our approach offers flexibility by distinguishing instances of the Topic class from Classes within domain ontologies, though mappings between them are possible. By avoiding the stringent logical constraints of a class-subclass hierarchy, we provide a more versatile way of annotating distributed content within KnowWhereGraph. Instead of building our topic hierarchy from scratch, we leveraged existing taxonomies and domain vocabularies developed by a wide range of specialists. We then arranged compound topics that span multiple domains within this preliminary hierarchy.


\subsection{Modeling Metadata for KnowWhereGraph}
\label{ssec:md-model}
Metadata plays a crucial role in KnowWhereGraph for multiple reasons: (1) KnowWhereGraph integrates data from heterogeneous sources; it is thus critical to understand the origin of the data; (2) Due to the size of KnowWhereGraph, It is important to maintain clear oversight of the graph’s contents in an easily queryable format; and (3) With a large and evolving team, it is thus important to document responsibility for maintenance and expertise purposes.

To effectively manage metadata and provenance of KnowWhereGraph, we adopt several established vocabularies. For instance, we utilize the Friend of a Friend Ontology (FOAF) \citep{foaf-tr} to represent the development team and their respective roles; the Dublin Core Metadata Initiative (DCMI) \citep{dcmi} Metadata Terms for attributes like title, creator, creation time, description, license, and right;  the PROV-O \citep{provo} to track the origin and lineage of resources, such as datasets; and the Simple Knowledge Organization System (SKOS) \citep{miles2009skos} for capturing definitions, the taxonomic hierarchy of domain concepts, and their examples.

\begin{figure}[t]
    \centering
    \includegraphics[width=\linewidth]{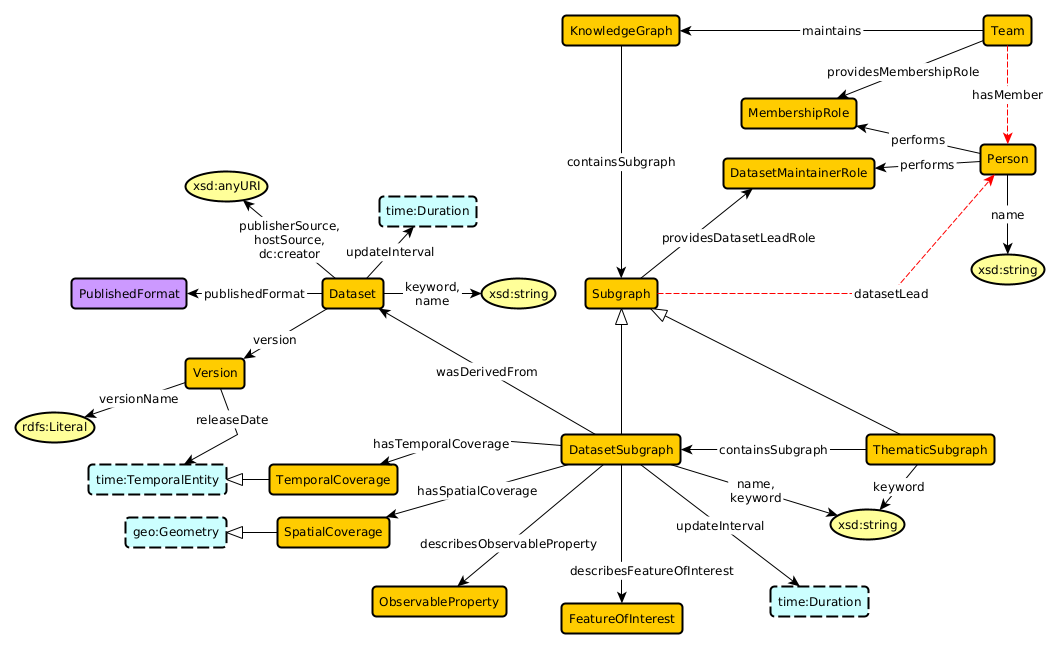}
    \caption{A schema diagram of KnowWhereGraph's metadata model, which includes a structured description of the KWG team and \textit{their roles} over time. The red dashed line indicates an axiomatically described shortcut.}
    \label{fig:md-model}
\end{figure}

Figure \ref{fig:md-model} shows a schema diagram for KnowWhereGraph's metadata model. The model captures (provenance) metadata at the \textsf{Dataset} level, \textsf{Subgraph} level, and for the whole KnowWhereGraph (\textsf{KnowledgeGraph} level). Essentially, datasets that are retrieved from different agencies---with different licensing---are materialized into a knowledge graph. Then, these knowledge graphs are integrated together into subgraphs (e.g., Expert-Expertise subgraph). These subgraphs collectively constitute KnowWhereGraph. For each of these different knowledge graphs, or collections thereof, maintainer roles are captured, with the duration of their validity. This allows KnowWhereGraph to answer such introspective questions as:
\begin{compactitem}
    \item Which team member $t$ materialized subgraph $s$?
    \item Where did dataset $d$ originate?
    \item How current is the materialized graph $g$?
\end{compactitem}
Furthermore, the metadata model hooks into the rest of the graph via observable properties and features of interest, to prevent ``triple-explosion'' \citep{triple-cost}. This allows us to capture or otherwise indicate a specific \textsf{Observation} origin from data taken from a specific agency, without having to directly specify this for every observation.

\section{Tools to Access KnowWhereGraph and Use Cases}
\label{sec:tools}
This section focuses on providing inter-disciplinary, cross-linked, and open-access geospatial data via KnowWhereGraph. The goal is to provide GIS users, as well as the general public who have no experience in Semantic Web technologies, the opportunity to use KnowWhereGraph. To do so, we first introduce two generalized tools developed to facilitate the access of KnowWhereGraph. We then present several use cases, as well as their customized tools, to demonstrate the potential of KnowWhereGraph in addressing specialized cross-disciplinary challenges. 

\subsection{Generalized Tools}
\subsubsection{Knowledge Explorer}
\begin{figure}[ht]
  \centering
  \includegraphics[width=\linewidth]{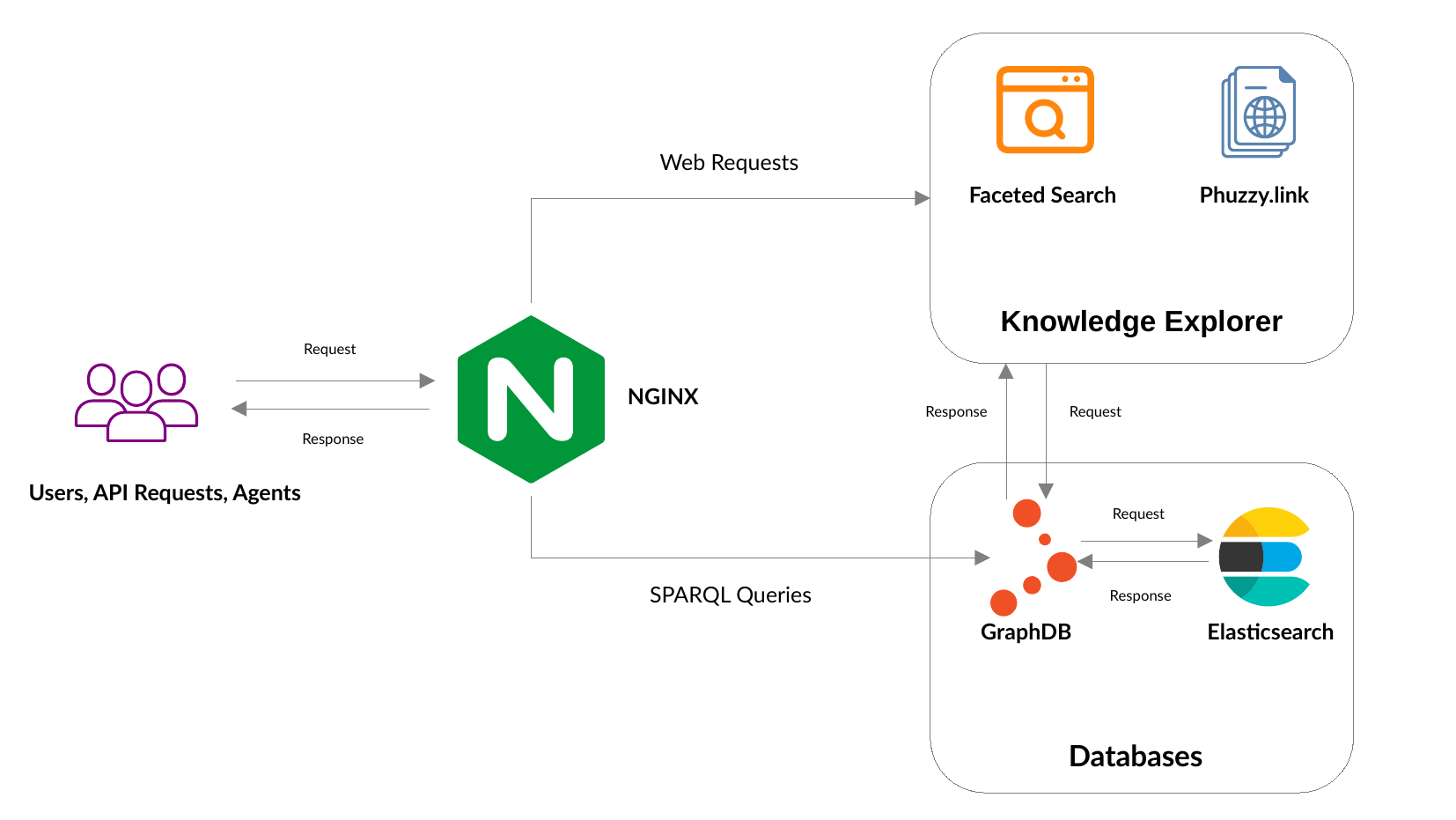}
  \caption{Architecture of Knowledge Explorer.}
\label{fig:architecture}
\end{figure}

Knowledge Explorer is a search engine that supports browsing the KnowWhereGraph
through predefined facets and a map interface \citep{liu2022knowledge}. Figure \ref{fig:architecture} illustrates its architecture design. Essentially, we adopt nginx, a Web server, to help manage
communications between end users, Knowledge Explorer, and KnowWhereGraph. Specifically, together with the dereferencing interface supported by
Phuzzy.link, a fully extensible Web interface to access RDF datasets \citep{regalia2017phuzzy}, the Knowledge Explorer provides a unified linked data approach for searching
based on predefined facets. To enable easy search
to end users, we also implement Elasticsearch\footnote{\url{https://www.elastic.co/elasticsearch}}, an open-source search framework, to
facilitate text-based queries to KnowWhereGraph.


Regarding the faceted search interface, we use two main classes defined in KnowWhereGraph ontology as the starting point: \textbf{Region} (labeled as ``Place" in the interface to make it user-friendly) and \textbf{Hazard} (see Section \ref{sec:schema}), because instances of these two classes are densely linked with others in the graph. More specifically, KnowWhereGraph is a \textit{gazetteer of gazetteers}. Its key strength is the structured geospatial backbone it provides, integrating geographic regions from heterogeneous data sources. Therefore, the ``Place" facet provides a way to search through several types of regions as modeled in Section \ref{ssec:st-region}, including \textit{Administrative Regions}, \textit{ZIP Code Area}, \textit{FIPS Code Area}, \textit{U.S. Climate Division}, and \textit{National Weather Zone}. Geographic feature types (implemented as facets too), such as schools, hospitals, and so forth, help to narrow the scope of search to particular feature types of interest. Moreover, as KnowWhereGraph is designed with environmental intelligence and humanitarian relief as the pilot use cases, Knowledge Explorer provides a facet on Hazard to improve its accessibility to users.  Based on user selection of hazard categories and spatiotemporal constraints, over two billion hazards related statements in the KnowWhereGraph can be retrieved so that the following example competency questions can be answered:
\begin{enumerate}
    \item What are the wildfires that crossed paths with an airport in Harris County, Texas?
    \item What are the official NOAA natural disasters that occurred in the East Carroll National Weather Zone over the last 20 years that had 5 or more associated deaths?
    \item What are the smoke plumes that have crossed over Salinas, California in the past three months?
\end{enumerate}



\begin{figure}[ht]
  \centering
  \includegraphics[width=\linewidth]{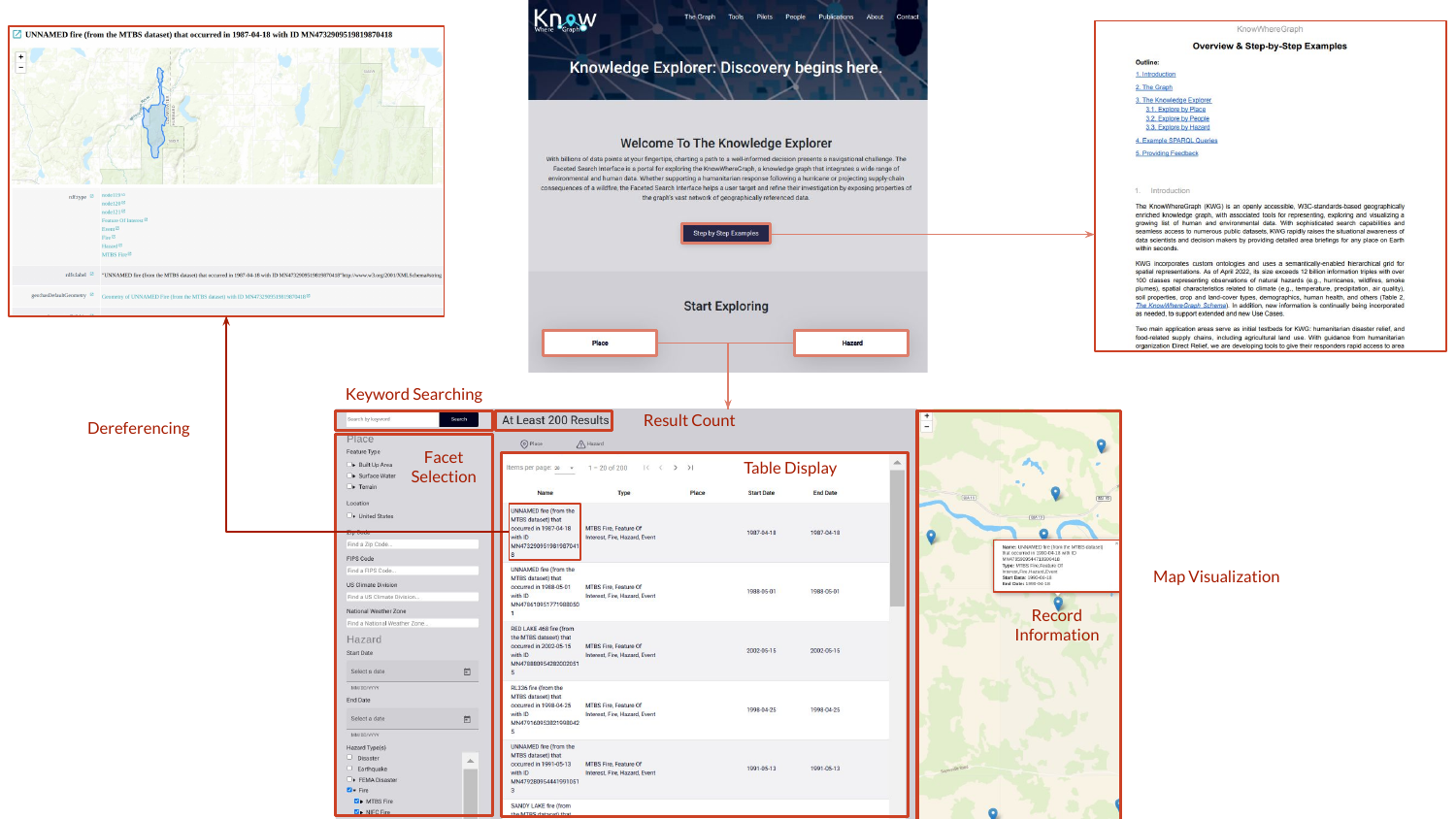}
  \caption{Knowledge Explorer in a nutshell. Top: Welcome Page; Bottom: The Faceted Search Interface. }
\label{fig:interface}
\end{figure}

Figure \ref{fig:interface} demonstrates different components of the Knowledge Explorer interface, including a Welcome Page with step-by-step instruction and starting facets, as well as the Faceted Search Interface with a left-hand panel on detailed facets, a middle panel showing data as a table-view, and a right-hand panel showing data as a map-view. 

\subsubsection{Geo-Enrichment Tools}
Geo-enrichment tools \citep{mai2019deeply,mai2022narrative} are a collection of GIS extensions to help connect popular GIS applications with geospatial knowledge graphs such as KnowWhereGraph. The goal is to enable GIS users to readily link multidisciplinary geospatial data, provided by KnowWhereGraph, into their research. To do so, we have developed Geo-Enrichment plugins for ArcGIS and QGIS in order to support both commercial and open-source GIS users.

\begin{figure}[ht]
  \centering
  \includegraphics[width=\linewidth]{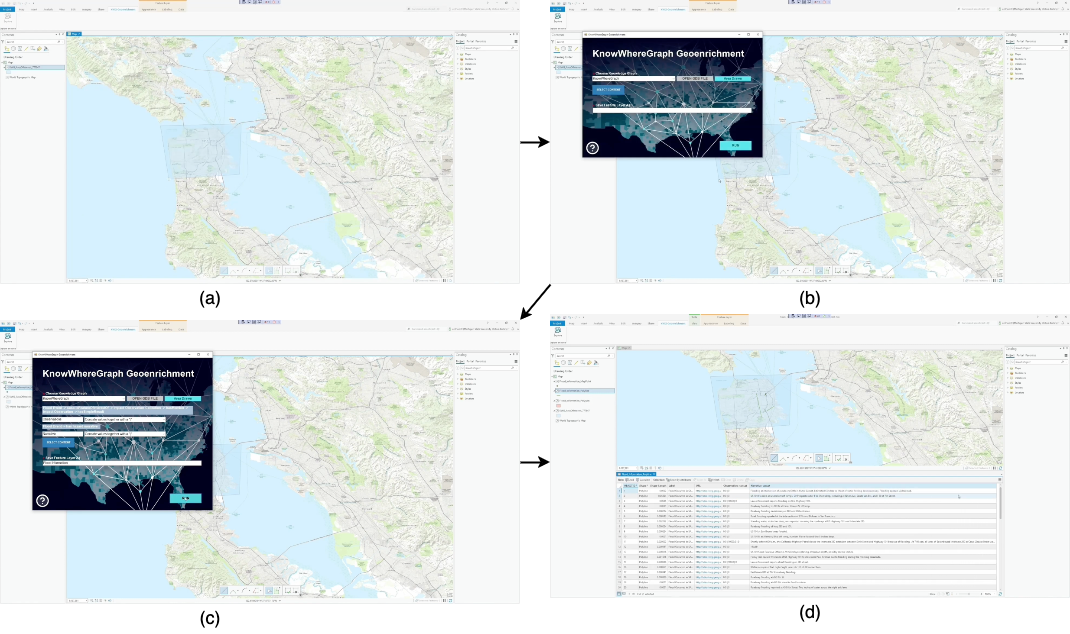}
  \caption{Demonstration of using Geo-Enrichment extension in ArcGIS to find flood events, and their related information, in a defined region surrounding San Francisco, U.S.: (a) To define a region of interest; (b) To connect with a knowledge graph (KnowWhereGraph by default); (c) To retrieve data from the graph by defining the chain of relations; (d) To generate the output as a shapefile with an attribute table.    }
\label{fig:geoenrichment}
\end{figure}

Figure \ref{fig:geoenrichment} provides an example of using the plugin in an ArcGIS Desktop environment. The guiding example question is to find all relevant information from KnowWhereGraph regarding those floods that have struck the region surrounding San Francisco. First, a user could define the study region by either importing a predefined geometry (e.g., as a shapefile) or drawing one on a map (as shown in Figure \ref{fig:geoenrichment}.a). With the region of interest being defined, a window will pop up for users to select which knowledge graph they would like to link to (Figure \ref{fig:geoenrichment}.b). Note here that by default, KnowWhereGraph is automatically provided as an option, while other open knowledge graphs, such as Wikidata and Yago\footnote{\url{https://yago-knowledge.org/}}, could be included too. Subsequently, users could select what kind of information will be retrieved from KnowWhereGraph. For instance, Figure \ref{fig:geoenrichment}.c demonstrates selecting observations of impact (i.e., the number of people who have died and/or injured?) and its narrative descriptions of any flood hazard that happened within the region of interest. These observations and associated attributes are automatically populated based on the ontology and data structure of KnowWhereGraph. In contrast to traditional GIS operations, such as spatial join, geo-enrichment enables the integration of data through ``follow-your-nose" semantics. Namely, one can retrieve all the information that is related to floods within the study region as long as they are connected in KnowWhereGraph directly or through multiple hops (i.e., a chain of relations). For instance, in addition to access attributes associated with the flood, users can retrieve hazards, as well as their attributes, that might have caused it, such as a wildfire that was observed nearby precedent the flood. Lastly, once the query chain is built, users can process it by generating the outputs, which are saved as shapefiles with different geometries (Figure \ref{fig:geoenrichment}.d). For instance, those floods with geometries of polyline will be saved as one file together with the retrieved attributes. If there were floods represented as points, a separate shapefile would be generated. 


\subsection{Customized Tools with Use Cases}
\subsubsection{Assessing Food Supply Chain Disruptions from Wildfire or Other Disasters}
Supply chain resilience assumes that considerable robustness and adaptability exist within the depth and overlap of diverse supply and demand networks. However, extreme weather events, such as wildfires and floods, inherently pose risks of network disruption and delayed recovery. In collaboration with our partner, the Food Industry Association (FMI)\footnote{\url{https://www.fmi.org/}}, we identified food quality and safety challenges, particularly those arising from natural disasters or disruptions, as a critical and urgent concern within the industry. 

The Wildfire Crop Impact Assessment tool is a front-end web interface designed to demonstrate the capabilities of KnowWhereGraph for our work with members of FMI. This tool is designed to walk decision-makers through a series of key questions that are key to assess the impact of active wildfires, associated smoke plumes, and their ash on the supply chain of fresh food products and retail operations. These questions are: (1) Where are/were fires happening in the U.S. for a given time period? (2) Are these fires causing smoke plumes, and what are their extents? (3) In areas of dense smoke, are there specific assets that may be affected? (4) Do measures on the ground, e.g., particulate matter (PM) pollution, indicate exposure of produce and workers? (5) What food products will be impacted by this hazard? (6) How will the food supply chain be impacted for a specific set of food producers, distributors, and retailers and how can the situation be remedied through alternate sourcing? FMI members currently carry out these analyses internally, but they are usually slow and backward-looking (i.e., not close enough to real-time) with limited data inputs, relying on available internal knowledge. 

\begin{figure}[ht]
  \centering
  \includegraphics[width=.9\linewidth]{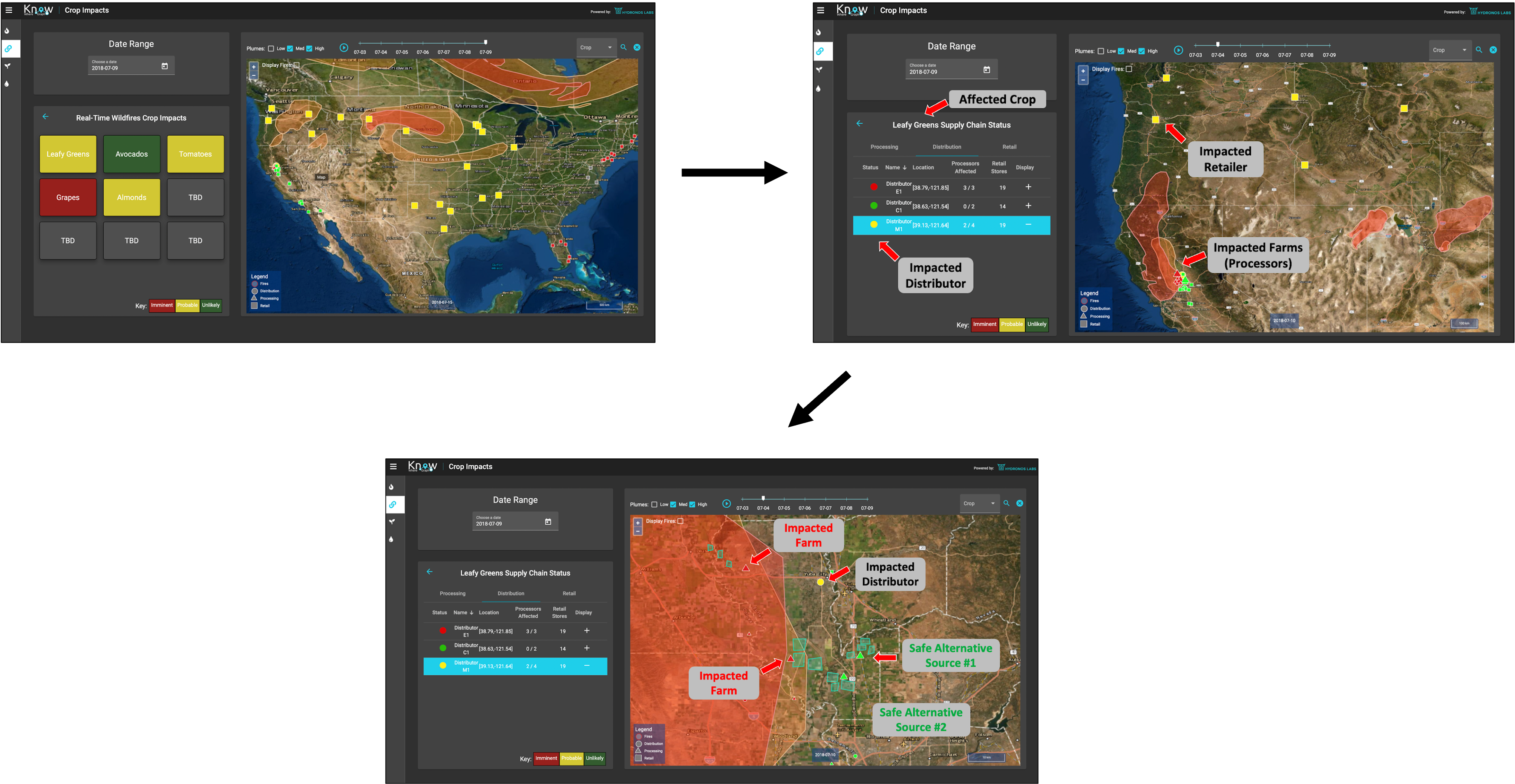}
  \caption{Demonstration of the Wildfire Crop Impact Assessment tool, illustrating the progression of user queries when identifying areas affected by wildfires and their supply chain impacts.}
\label{fig:supplychaininterface}
\end{figure}

As illustrated in Figure \ref{fig:supplychaininterface}, users can navigate through these aforementioned questions without any prior experience with complex GIS software, the specialized data, or advanced analytical methods involved. The results are visualized at each step of the process. Essentially, the tool is programmed to automatically construct and send (Geo)SPARQL queries to KnowWhereGraph given users' input, such as the selection of data layers, and drawing a study region on the map, etc.  Responses received from the graph are further processed and visualized in the tool with a user-friendly interface, allowing users to effectively understand relevant situations surrounding the crop lands that are selected. 


\subsubsection{Land Potential Assessment for Valuation and Risk Assessment}


In collaboration with our partner, Farm Credit Association, we explored the use of KnowWhereGraph to assess land potential—an essential factor in evaluating loan application risks for farmers, particularly after environmental disasters. The complexity of the agricultural credit system, which includes appraisals from local experts, crop yield projections under various environmental scenarios, and prepayment forecasts, requires a diverse range of data. This includes information on farm equipment, farming practices, local weather conditions, crop types, and projected crop prices. KnowWhereGraph provides a centralized platform for stakeholders to access and analyze these data efficiently in one place.   

\begin{figure}[ht]
  \centering
  \includegraphics[width=.9\linewidth]{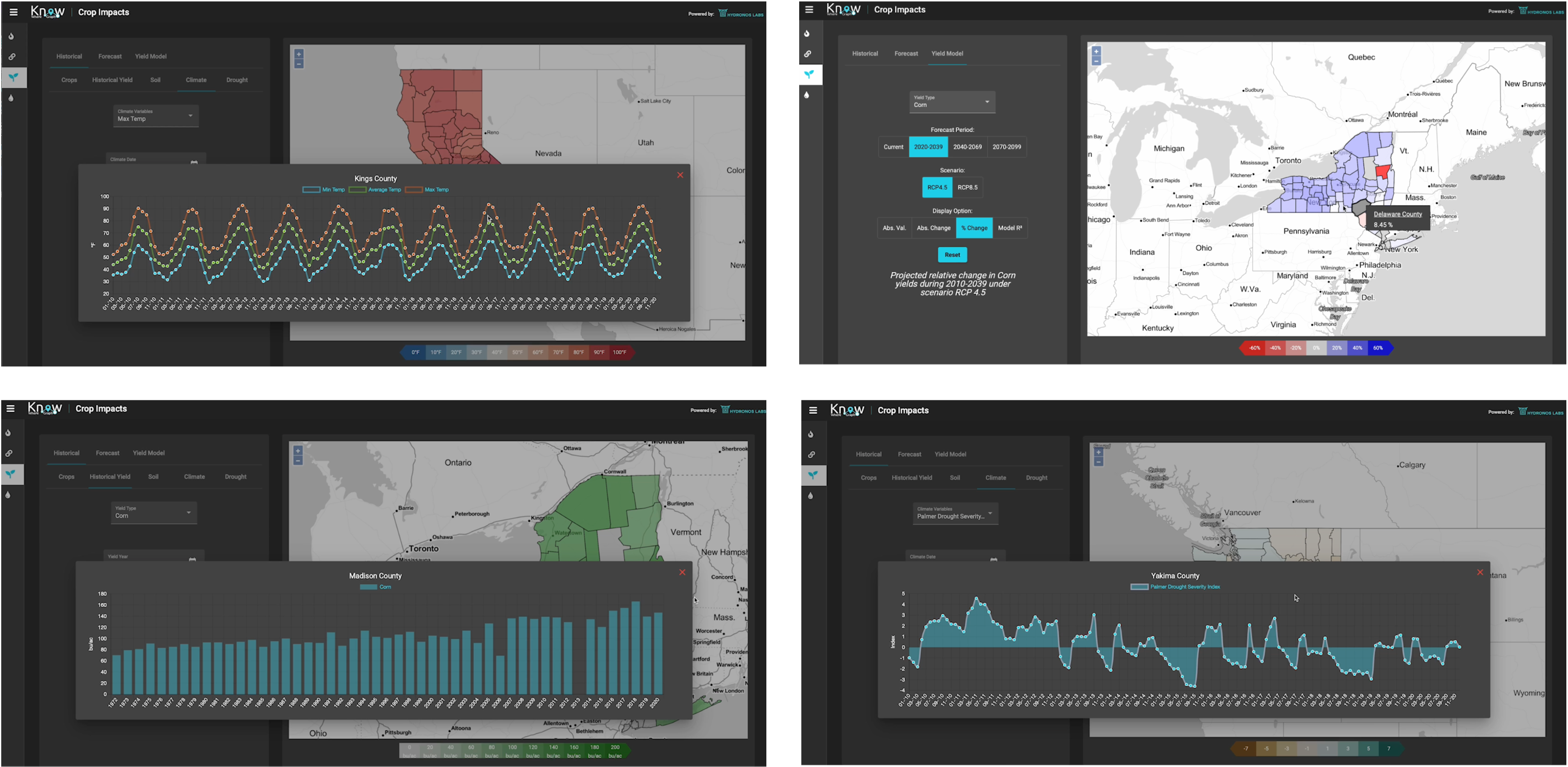}
  \caption{Sample visualizations from the custom applications developed for Farm Credit Association internal use to evaluate the impact of climate and weather events on land valuation. KnowWhereGraph enables exploratory data analysis and modeling through these interfaces.}
\label{fig:farmcreditinterface}
\end{figure}

Ultimately, we illustrated the enhanced analytical powers of knowledge graph technology and KnowWhereGraph through a prototype Web application investigating climate change and risk drivers for selected crops in three designated regions: San Joaquin Valley, Central/Western New York, and Columbia Basin. This tool, shown in Figure \ref{fig:farmcreditinterface}, enables end users to query the graph and answer a series of specific workflow questions, all without the need for specific data or scientific expertise, including:

\begin{enumerate}
    \item Where are the crops of interest being grown in the U.S.? Where have they been grown in the past?
    \item What are the physical (soil) properties of this agricultural land?
    \item How has the current (and past) climate affected productivity on this land?
    \item How will the region of interest be affected by a changing climate?
    \item How will this impact the production of specific crops and how will this ultimately affect the value of the land?
\end{enumerate}

Farm Credit Association representatives were encouraged to use this tool and provide feedback for further development, with the goal of ultimately building a KnowWhereGraph-based tool to inform decision-making in the sector. 

\subsubsection{Disaster Response and Humanitarian Aid from Hurricane Disasters}
GeoGraphVis is a visualization tool that supports disaster response and humanitarian aid through the integration of KnowWhereGraph \citep{li2023geographvis}. Given the increasing frequency and intensity of disasters in the past decade, such as hurricanes, floods, and wildfires, the need for timely and effective disaster management has become more urgent. GeoGraphVis addresses this challenge by enabling relief experts to quickly acquire and visualize relevant contextual information about affected areas, infrastructures, and populations, so as to support evidence-based decision-making in response to disasters. 

\begin{figure}
  \centering
  \includegraphics[width=.9\linewidth]{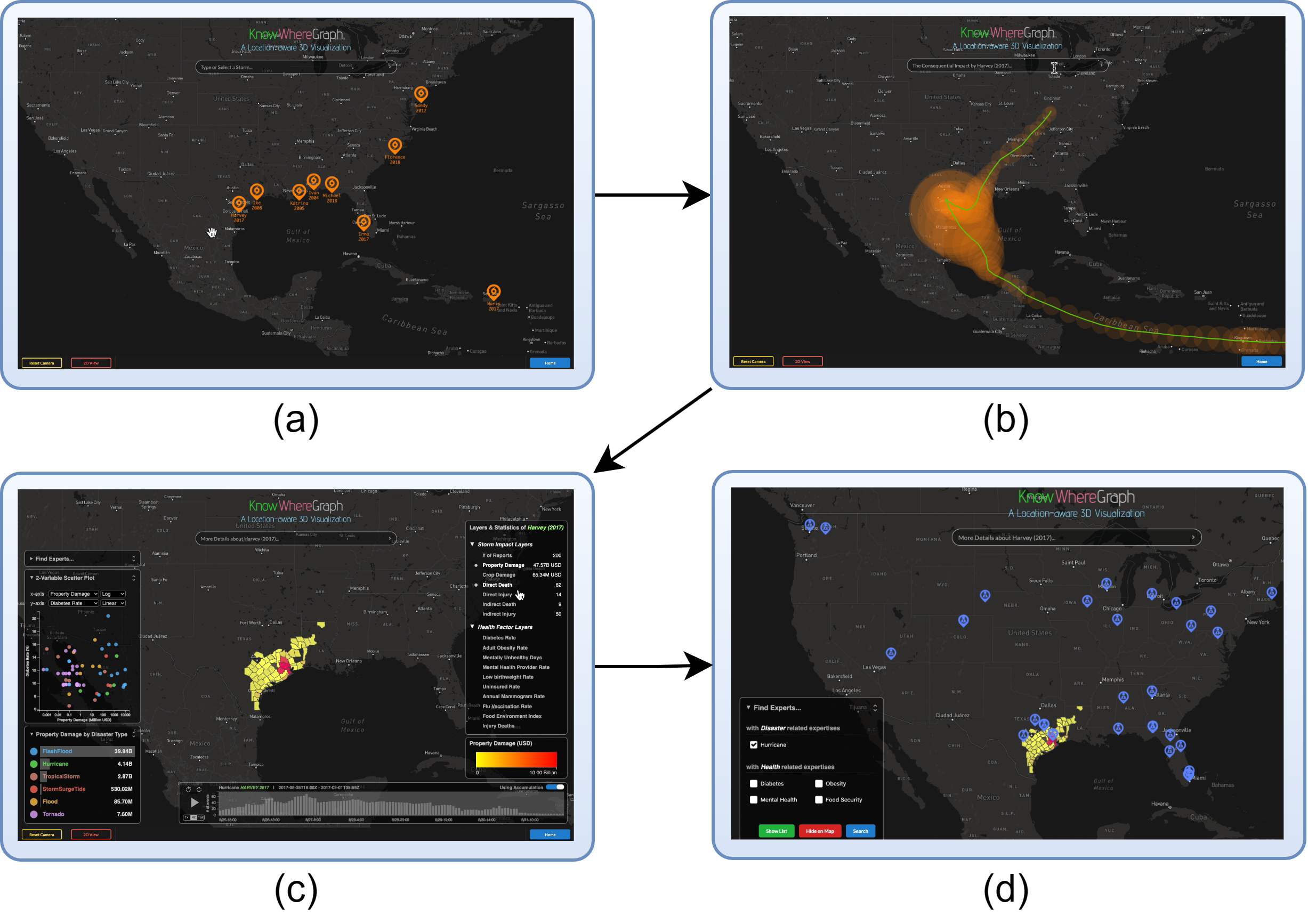}
  \caption{The overall workflow of GeoGraphVis. Each subgraph shows (a) a storm list and the most impactful hurricanes in history; (b) the trajectory of a selected hurricane; (c) hurricane damage and health profile of the impact area of the selected hurricane; (d) the distribution of experts with desired expertise.}
  \label{fig:geographvis}
\end{figure}

GeoGraphVis leverages KnowWhereGraph to provide a range of functionalities critical for disaster management. The overall workflow is illustrated in Figure \ref{fig:geographvis}. This tool enables users to query specific disasters (e.g., hurricanes) and retrieve detailed spatio-temporal data (e.g., trajectories, wind speed, and other indicators). The visualization employs choropleth maps to represent various statistics describing disasters, such as the extent of property damage across the affected regions, with color gradations to indicate severity \citep{maps}. This function allows users to quickly assess the impact on infrastructure and population as well as identify areas requiring immediate attention.

In addition to visualizing damage, GeoGraphVis displays health data (e.g., diabetes and obesity rates) from KnowWhereGraph at the county level to help identify vulnerable populations. This multifaceted tool enhances the understanding of the disaster's impact by allowing users to correlate health vulnerabilities with the disaster's damage. It features interactive scatter plots that enable the simultaneous analysis of multiple variables from KnowWhereGraph. In this way, it helps users to identify which counties are most affected by both disasters and health risks; for instance, it visualizes property damage in relation to diabetes rates to pinpoint more vulnerable communities (that may need additional support).

Another feature GeoGraphVis supports is mapping experts and their relevant expertise from KnowWhereGraph (see Section \ref{ssec:expert}), providing users with immediate access to specialized knowledge during disaster response. Users can quickly find and connect with local experts (e.g., those in the field of disaster management, public health, and environmental science) by simply visualizing their locations on the same geospatial interface. This integration further fosters collaboration (e.g., between researchers and decision-makers) and ensures the decision-makers have the relevant expertise needed during a disaster event.


\section{Conclusion and Future Work}
\label{sec:conc}
Integrating geographic data so that they can be effectively discovered, queried, analyzed, reasoned, and visualized is one of the fundamental goals of developing GI Systems and GI Science. In contrast to the traditional way of managing geographic data as tables and layers, KnowWhereGraph proposes a new data integration and management framework at the intersection of human and environmental systems. This framework harmonizes various types of geographic data, including observations in distinct modes (e.g., maps, remotely sensed images), attributes, spatial and temporal references, data scheme, and metadata, in a homogeneous environment using a labeled directed multi-graph, where both nodes and edges are semantically annotated via formalized ontologies. By doing so, KnowWhereGraph delivers a densely interconnected (i.e., pre-integrated) and AI-ready (i.e., human- and machine-readable and reason-able) data structure ready for use in systems such as retrieval-augmented generation (RAG) \citep{edge2024local,zhou2024img2loc,yu2025spatial} powering many modern AI systems. In such a way, KnowWhereGraph also fosters cross-disciplinary knowledge discovery and collaboration as well as geo-enrichment tasks across domains. This paper introduced the design principle of KnowWhereGraph, its core schema and data sources, its innovations, tooling, and multiple use cases. Notably, in contrast to other large-scale knowledge graphs, the KnowWhereGraph leverages space and time as first-class citizens, i.e., as the \textit{nexuses} to interlink knowledge from different disciplines, emphasizing the role geography plays in cross-disciplinary research. In fact, we see KnowWhereGraph as a gazetteer of gazetteers \citep{janowicz2022know,frew1998alexandria} that provides different types of geographic identifiers from named geographic places to hierarchical S2 cells, but in contrast to classical gazetteers, KWG delivers deep (meta)data about each of these places and regions. Other graphs can then topologically link to each other by linking into KnowWhereGraph.

Thanks to its extensibility, KnowWhereGraph continues to evolve, incorporating new schemas and datasets based on emerging use cases. A key priority is to expand the graph from a primarily U.S.-focused database to a global resource. This expansion has the potential to support domain scientists and decision-makers worldwide in addressing large-scale, global challenges, such as public health crises and climate change. Currently, KnowWhereGraph focuses on pilot use cases related to supply chain management, disaster response, humanitarian relief, and land value assessment. However, as the work develops, we aim to broaden its reach to a wider community, including government agencies, academic institutions, NGOs, and the third sector. This will help empower diverse stakeholders to leverage KnowWhereGraph for evidence-informed decision-making. Finally, as one of the largest knowledge graphs focused on geographic data, KnowWhereGraph offers a unique opportunity to align AI methodologies with real-world societal and environmental applications. For instance, it would be valuable to explore using KnowWhereGraph as contextual knowledge to enhance “black-box” generative AI models, such as large language models, thereby improving the credibility and reliability of AI-driven responses.




\section*{Data and code availability}
The data and code developed in this project is shared on Github with the link: \url{https://github.com/KnowWhereGraph }

\section*{Funding}
We acknowledge funding by the US National Science Foundation Award "A1: KnowWhereGraph: Enriching and Linking Cross-Domain Knowledge Graphs using Spatially-Explicit AI Technologies" (\#2033521).

\section*{Acknowledgments}

We dedicate this paper to our project member Dr. E. Lynn Usery of the U.S. Geological Survey and our steering committee member Prof. Peter Fox of Rensselaer Polytechnic Institute, who passed away on March 22, 2022 and March 27, 2021, respectively. Both have significantly contributed to the KnowWhereGraph project and the entire geo-semantics community.



\bibliographystyle{plainnat}
\bibliography{refs}
\clearpage
\appendix
\section{Agency Acronyms}
\label{app:acronym}
\begin{table*}[h!]
  \label{tab:acronyms}
  \centering \small \setlength{\tabcolsep}{8pt}
  \makebox[\textwidth]{\begin{tabular}{ll}
    \toprule
    Acronym & Full Name \\
    \midrule
    CDC     & Centers for Disease Control and Prevention \\
    FEMA    & Federal Emergency Management Agency \\
    HRSA    & Health Resources and Services Administration \\
    MTBS    & Monitoring Trends in Burn Severity \\
    NACCHO  & National Association of County and City Health Officials \\
    NDMC    & National Drought Mitigation Center \\
    NIFC    & National Interagency Fire Center \\
    NOAA    & National Oceanic and Atmospheric Administration \\
    NRCS    & Natural Resources Conservation Service \\
    UN OCHA & United Nations Office for the Coordination of Humanitarian Affairs \\
    USCB    & United States Census Bureau \\
    USDA    & United States Department of Agriculture \\
    USDHS   & United States Department of Homeland Security \\
    USDOJ   & United States Department of Justice \\
    USDOT   & United States Department of Transportation \\
    USEPA   & United States Environmental Protection Agency \\
    USFS    & United States Forest Service \\
    USGS    & United States Geological Survey \\
    \bottomrule
  \end{tabular}}
\end{table*}
\end{document}